\documentclass[onecolumn,
preprintnumbers,
nofootinbib,
amsmath,amssymb,
aps,
prd,
]{revtex4-2}

\usepackage{graphicx}
\usepackage{dcolumn}
\usepackage{bm}
\usepackage{hyperref}
\usepackage{multirow}
\usepackage{slashed}
\usepackage{placeins}
\DeclareUnicodeCharacter{2212}{\textendash}
\def\be{\begin{equation}}
\def\ee{\end{equation}}
\def\bea{\begin{eqnarray}}
\def\eea{\end{eqnarray}}

\usepackage{makecell}

\newcommand{\nn}{\nonumber}

\usepackage{xcolor}

\makeatletter
\newcommand{\Biggg}{\bBigg@{3.5}}
\makeatother

\begin{document}

\begin{flushright}
TUM-HEP-1585/25\\
NORDITA 2025-168\\
\vspace*{1cm}
\end{flushright}

\title{Dark Acoustic Oscillations as an Early-Universe Explanation of the DESI Anomaly}

\author{Mathias Garny$^1$}\email{mathias.garny@tum.de}
\author{Florian Niedermann${}^2$}\email{florian.niedermann@su.se}
\author{Martin S.~Sloth${}^3$}\email{sloth@sdu.dk}
\affiliation{%
\small ${}^1$Physik Department T31, School of Natural Sciences, Technische Universit\"at M\"unchen \\
James-Franck-Stra\ss e 1, D-85748 Garching, Germany\\
${}^2$Nordita, KTH Royal Institute of Technology and Stockholm University\\
Hannes Alfv\'ens v\"ag 12, SE-106 91 Stockholm, Sweden\\
${}^3$ Universe-Origins, University of Southern Denmark, Campusvej 55, 5230 Odense M, Denmark}%

\begin{abstract}
DESI DR2 data have been widely interpreted as evidence for late-time evolving dark energy (DE) with an apparent phantom crossing. Here we investigate an alternative explanation, based on early-Universe physics. If dark acoustic oscillations (DAO) are close in scale to  baryon acoustic oscillations (BAO), they can bias the extraction of the BAO scale from the peak in the galaxy correlation function. This leads to an apparent shift in the inferred distance if the superposition of BAO and DAO features is misinterpreted as being due to BAO only. Taking this shift into account, we find that a DAO with percent-level amplitude can reconcile DESI DR2 with Planck 2018 as well as Pantheon+ supernovae data, with fit improvement at a similar level compared to evolving DE. Notably, a DAO feature with the required properties has been predicted in a previously proposed scenario that resolves the Hubble tension via a pre-recombination decoupling of dark matter and dark radiation (DRMD). The presence of a DAO feature close to the BAO peak can be scrutinized with future full-shape galaxy clustering data from DESI and Euclid.
\end{abstract}

\maketitle

\clearpage

\section*{Introduction}

The recent DESI DR2 data indicate evidence against the $\Lambda$CDM model. In a combined fit to  cosmic microwave background (CMB) Planck 2018 data,  DESI DR2 baryon acoustic oscillation (BAO) and Pantheon+ supernovae (SN) measurements, the DESI collaboration finds $2.8\sigma$ evidence for dynamical dark energy (DE) with $w_0\neq -1$, $w_a \neq 0$, where $w_0=-1$ and $w_a=0$ is the prediction of non-dynamical dark energy described by a cosmological constant in the $\Lambda$CDM model \cite{DESI:2025zgx}.

While these results are exciting, their interpretation in terms of dynamical DE also gives rise to some reasons for concern.
Assuming the values $w_0 =-0.838 \pm 0.055$, $w_a =-0.62^{+0.22}_{-0.19}$ preferred by DESI DR2 combined with Planck 2018 and Pantheon+ would imply that DE crossed into a phantom regime $w(z)<-1$ in the past, while the Universe was still DE dominated. This raises some immediate worries about the physical interpretation. A single field minimal quintessence model has an equation of state $w(z) \geq -1$, and  has been found to not be favored by the DESI DR2 result, see e.g.~\cite{Bayat:2025xfr},
although there is a possibility that a quintessence model with non-trivial dark matter (DM) interactions could mimic the phantom behavior~\cite{Khoury:2025txd,Kou:2025yfr,Chen:2025ywv,Liu:2025bss,Wang:2025znm,Caldwell:2025inn,SanchezLopez:2025uzw,Bedroya:2025fwh}. Late-time quintessence models (with or without DE-DM interactions) would also not, in their simplest form, be able to address other important tensions in the data, such as the Hubble tension~\cite{H0DN:2025lyy} and the CMB-BAO tension \cite{Ferreira:2025lrd}, as these tensions are known to require new early-time physics, before recombination \cite{Bernal:2016gxb,Knox:2019rjx}. 

Here we  consider an alternative interpretation of the DESI DR2 result, motivated by a recent proposed solution to the Hubble tension and the CMB-BAO tension based on new early-time dark sector physics. This opens up the possibility that the DESI DR2 anomaly is not a manifestation of dynamical DE, but instead another face of the same physics beyond $\Lambda$CDM manifesting itself also in the Hubble tension. In the Hot New Early Dark Energy (Hot NEDE) model of \cite{Niedermann:2021vgd,Niedermann:2021ijp,Cruz:2023lnq,Garny:2024ums,Garny:2025kqj}, a key feature in its success in solving the Hubble tension is the Dark Radiation-Matter Decoupling (DRMD) \cite{Garny:2025kqj,Garny:2024ums}, which also predicts the presence of a dark acoustic oscillation (DAO) \cite{Garny:2025kqj}, a smoking gun, which can be used to test the model and discriminate it from other proposed solutions to the Hubble tension, like axiEDE \cite{Poulin:2018cxd}, Cold NEDE \cite{Niedermann:2019olb,Niedermann:2020dwg,Cruz:2023lmn}, or varying electron mass with spatial curvature \cite{Schoneberg:2024ynd,Poulin:2024ken}. 
An additional component of self-interacting dark radiation in the early Universe has been proposed as a solution to the Hubble tension already early on \cite{Jeong:2013eza,Buen-Abad:2015ova,Bernal:2016gxb,Buen-Abad:2017gxg,Archidiacono:2020yey,Blinov:2020hmc,Aloni:2021eaq}. It does not provide a full solution to the Hubble tension, but only partially reduces it. However, the main obstacle is that the required amount of dark radiation is ruled out by Big Bang Nucleosynthesis (BBN) constraints \cite{Schoneberg:2022grr,Garny:2024ums}. 

In \cite{Garny:2025kqj,Garny:2024ums} it was proposed that a thermal bath of self-interacting dark radiation is created in the epoch between BBN and recombination by a dark $\mathrm{SU}(N)$ gauge symmetry broken to $\mathrm{SU}(N-1)$ in a supercooled phase transition. This avoids the BBN constraints on $\Delta N_\text{eff}$ that rule out other related scenarios of self-interacting dark radiation. In addition, in this setup a fermion SU(N) multiplet provides a  stable component of dark matter with gauge interactions with the self-interacting dark radiation consisting of non-Abelian gauge bosons. The spontaneous symmetry breaking predicts a decoupling of the interaction between dark matter and dark radiation once the dark sector cools down, which is referred to as dark radiation matter decoupling (DRMD). In comparison to other scenarios of self-interacting dark radiation, the Hot NEDE model with DRMD provides a full resolution to the Hubble tension when comparing the standard datasets of CMB (Planck 2018), BAO (DESI DR2), and SN (Pantheon+) data to the SH$_0$ES measurement of the local $H_0$ value. The Hot NEDE model with DRMD therefore stands out from generic models of self-interacting dark radiation in two ways: by being consistent with BBN data and by fully resolving the Hubble tension within the above-mentioned standard datasets \cite{Garny:2025kqj}. 

The DRMD is a dark parallel to the usual decoupling of photons at the time of recombination in the visible sector. While recombination and the subsequent decoupling is imprinted both in the CMB and in the BAO feature of Large Scale Structure (LSS), the DRMD causes similar acoustic oscillations in the dark plasma, to be imprinted in the DM and thus the LSS as a dark acoustic oscillation (DAO) \cite{Garny:2025kqj}, while not having a direct imprint on the CMB. The DAO is predicted to be similar in amplitude and close to the BAO, which, as we will show below, would make it easy for DESI to confuse the actual BAO with the DAO, inducing a linear DAO bias in the BAO measurement, obscuring cosmological parameter inference and leading to misleading conclusions. There are related scenarios predicting a DAO feature,  such as  Atomic Dark Matter~\cite{Kaplan:2009de,Cyr-Racine:2012tfp,Cyr-Racine:2021oal,Blinov:2021mdk,Bansal:2022qbi}. Dark atoms can undergo recombination similar to visible sector atoms, leading to a DAO feature in the matter power spectrum, and allowing for larger values of $H_0$~\cite{Buen-Abad:2024tlb,Buen-Abad:2025bgd,Bansal:2021dfh}, while satisfying the BBN constraint on extra radiation would require additional assumptions in this setup.

To remain as model independent as possible, but motivated by the above predictions, we assume here a generic DAO similar in amplitude and close to the BAO, and model the DAO bias producing a linear shift in the observed BAO data. We use this to constrain the DAO feature and infer the true constraints on cosmological parameters in the presence of a DAO. 
Our study serves as a motivation for further investigation of the DAO feature in model specific setups, like the DRMD model, and as an indication that the DESI anomaly could be hinting at something else than late-time evolving dark energy, but just as profound. 

There are two physically distinct configurations for the position of the DAO peak in the galaxy pair correlation function. When the DAO is almost on top of the BAO but at slightly larger scale, it shifts the perceived peak of the BAO in small positive direction of the DAO; we call this the positive branch. If the BAO and the DAO peaks are well separated but close, the characteristic tail of the DAO can shift the perceived position of the BAO in the opposite direction away from the DAO, which we call the negative branch. In this case the main peak of the DAO is assumed to be at a sufficiently small scale where it could be hidden by non-linear systematics in currently publicly available DESI DR2 data. The viability of the latter scenario is arguably more speculative, requiring further follow-up study. Both the positive and negative branches could potentially be discoverable in future DESI data releases or upcoming surveys like Euclid, by either resolving the DAO peak that overlaps closely with the BAO, or identifying a shallow DAO peak on scales $\sim 50$Mpc$/h$ somewhat smaller than the BAO. We note that the DAO considered here is well consistent with existing searches for extra features based on BOSS data~\cite{Beutler:2019ojk}, which weaken if the DAO is close in scale to the BAO.

The fit improves by $\Delta \chi^2 \simeq -5$ for both the positive and negative branch, if we include such a DAO effect in our combined analysis with DESI DR2, Planck 2018 CMB, and Pantheon+ SN data.
On the positive branch, we find  a viable scenario with $\Delta r/r_B\simeq 0.2$, where $r_B$ is the position of the BAO peak, and $\Delta r\equiv r_D -r_B$, with $r_D$ being the position of the DAO peak, as well as relative amplitude $\tilde A=A_D/A_B\simeq 0.2$ of DAO and BAO features in the linear power spectrum, corresponding to a percent-level absolute amplitude $A_D\simeq 0.01$. On the negative branch, we find that the relative position of the DAO peak compared to the position of the BAO peak preferred by DESI DR2 data is  $\Delta r/r_B \in [-0.53,-0.34] $ (68\% CL). In the Hot NEDE model with DRMD\footnote{And related models addressing the Hubble tension through a decoupling of dark radiation and dark matter, like Atomic Dark Matter \cite{Buen-Abad:2024tlb,Buen-Abad:2025bgd,Bansal:2021dfh}.}, the independent prediction, driven by the model's ability to address the Hubble tension is $\Delta r/r_B \in [-0.48,-0.32] $ (68\% CL)~\cite{Garny:2025kqj}. While, as mentioned before, the viability of the negative branch scenario needs to be further investigated, it is intriguing that the position of the DAO hinted at by DESI matches closely with the previously found prediction of the DRMD model.

The fit improvement obtained for a DAO is at a similar level as for $w_0\neq -1$, $w_a \neq 0$ and no DAO for the same data set combination ($\Delta \chi^2 \simeq -5$ vs $\Delta \chi^2 \simeq -9$), making a DAO an early-Universe alternative to evolving DE as an explanation to the DESI DR2 anomaly.
While other alternative explanations of dynamical DE have been discussed, they consider physics affecting the late Universe evolution. In addition, some of these alternatives, such as allowing for spatial curvature~\cite{Chen:2025mlf} or a modified opacity from reionization~\cite{Sailer:2025lxj}, reconcile only CMB and BAO distance data but cannot account for SN measurements at the same time, in contrast to the DAO scenario considered here.

Below we discuss how we model the impact of a DAO close to the BAO, which can be implemented into the DESI DR2 data analysis as a systematic bias of the inferred BAO scale. Then we discuss our data analysis and results, and after that we conclude. 

\section*{Dark acoustic oscillation systematics}

We want to estimate the possible effect of DAO on the BAO measurements of DESI. 
Let us first briefly review how the BAO peak location is extracted from the measured galaxy pair-correlation function by the DESI collaboration~\cite{DESI:2025qqy}, following the setup described in~\cite{Chen:2024tfp}.
The BAO feature encoded in galaxy pairs along the line of sight is sensitive to $D_H(z)/r_{d,B}$, and to $D_M(z)/r_{d,B}$ in the perpendicular directions, with $D_H(z)=c/H(z)$ and angular distance $D_M(z)=c\int_0^z dz'/H(z')$ for a flat Universe as well as BAO drag horizon $r_{d,B}$ and Hubble rate $H(z)$. We introduce the subscript to discriminate the BAO scale from the DAO considered below, and adopt the notation $r_B\equiv r_{d,B}$ for simplicity. The values of the measured BAO distance parameters are reported relative to a fiducial reference cosmology~\cite{DESI:2025qqy} (a $\Lambda$CDM model with parameters close to the best fit for Planck 2018), via
\be
  \alpha_\parallel(z) \equiv \frac{D_H(z)/r_{B}}{(D_H(z)/r_{B})^\text{fid}},\qquad
  \alpha_\perp(z) \equiv \frac{D_M(z)/r_{B}}{(D_M(z)/r_{B})^\text{fid}}\,,
\ee
and the combination $\alpha_\text{iso}(z)\equiv (\alpha_\parallel\alpha_\perp^2)^{1/3}$ is related to the angle-averaged BAO distance scale $D_V(z)\equiv (zD_H(z)D_M(z)^2)^{1/3}$.

The extraction of the BAO peak location in the galaxy pair correlation function relies on a decomposition of the underlying power spectrum into a smooth broadband component and an oscillatory contribution featuring a damping due to decorrelation by a stochastic ensemble of large-scale bulk flows with a strength that depends on $\mu$, the cosine of the angle between the wave vector and the line of sight. The technique of BAO reconstruction attempts to (partially) undo this decorrelation from the measured large-scale density field. For both the case with and without applying BAO reconstruction, the galaxy power spectrum is modeled as~\cite{DESI:2025qqy}
\be\label{eq:Pgmodel}
  P_g(k,\mu) = B(k,\mu)P_{\text{nw}}(k)+C(k,\mu)P_{\text{w}}(k)+D(k)\,,
\ee
where $P_{\text{nw}}(k)$ and $P_{\text{w}}(k)$ stand for the smooth and oscillating contributions to the linear matter power spectrum $P_{\text{lin}}(k)=P_{\text{nw}}(k)+P_{\text{w}}(k)$, and the terms involving $B(k,\mu)$ as well as $D(k)$ describe the broadband contribution as given in Eqs.~(15) and (17) in~\cite{DESI:2025qqy}, following~\cite{Chen:2024tfp}. The BAO peak is encapsulated in the term involving $P_{\text{w}}(k)$, with [see Eq.~(16) in~\cite{DESI:2025qqy}]
\be
  C(k,\mu)=(b_1+f\mu^2)^2\exp\left[-\frac12 k^2 \left( \mu^2\Sigma_\parallel^2 + (1-\mu^2)\Sigma_\perp^2 \right)\right]\,,
\ee
where the first term contains the redshift-space distortion in Kaiser approximation with linear bias $b_1$ and growth rate $f$ (see Table II in~\cite{DESI:2025qqy} for the values adopted for the various galaxy populations measured by DESI), while the exponential captures the damping of the BAO in Gaussian approximation~\cite{Eisenstein:2006nj,Seo:2007ns}, with damping scales $\Sigma_{\parallel,\perp}$ along and across the line-of-sight direction, respectively. While the DESI analysis treats both of these as free parameters, with fiducial values and prior ranges reported in Table III in~\cite{DESI:2025qqy}, they have also been computed by resumming those loop contributions to the power spectrum that are enhanced for  loop wave number in the IR range $1/r_B\lesssim q\lesssim k$~\cite{Matsubara:2007wj,Matsubara:2008wx,Carlson:2012bu,Senatore:2014via,Baldauf:2015xfa,Seo:2015eyw,Vlah:2015zda,Blas:2016sfa,Ivanov:2018gjr}. The power spectrum is decomposed into multipole moments, and subsequently Fourier transformed to obtain the correlation functions
\be\label{eq:xiell}
\xi_\ell(s)=\frac{i^\ell}{2\pi^2}\int_0^\infty dk k^2 j_\ell(ks)P_\ell(k),\qquad  P_\ell(k)=\frac{2\ell+1}{2}\int_{-1}^1 d\mu\, P_g(k',\mu') {\cal L}_\ell(\mu)\,,
\ee
where $j_\ell$ and ${\cal L}_\ell$ are spherical Bessel functions and Legendre polynomials, respectively. Here $P_g(k',\mu')$ is evaluated for the fiducial cosmology, but with rescaled wave number along and perpendicular to the line of sight according to $k_\parallel'=k_\parallel/\alpha_\parallel$ and $k_\perp'=k_\perp/\alpha_\perp$, where $k_\parallel=k\mu, k_\perp=k\sqrt{1-\mu^2}$, and analogously for the primed quantities, as well as $\mu' = [1+(k_\perp \alpha_\parallel)^2/(k_\parallel \alpha_\perp)^2]^{-1/2}$. The measured values for $\alpha_\parallel(z_i)$ and $\alpha_\perp(z_i)$ are obtained by fitting this theory model to the measured correlation functions $\xi_\ell(s)$ for the monopole and quadrupole $\ell=0,2$, and each galaxy sample at effective redshifts $z_i$.

\begin{figure}[t]
\centering
\begin{minipage}{0.5\textwidth}
    \centering
    \includegraphics[width=\textwidth]{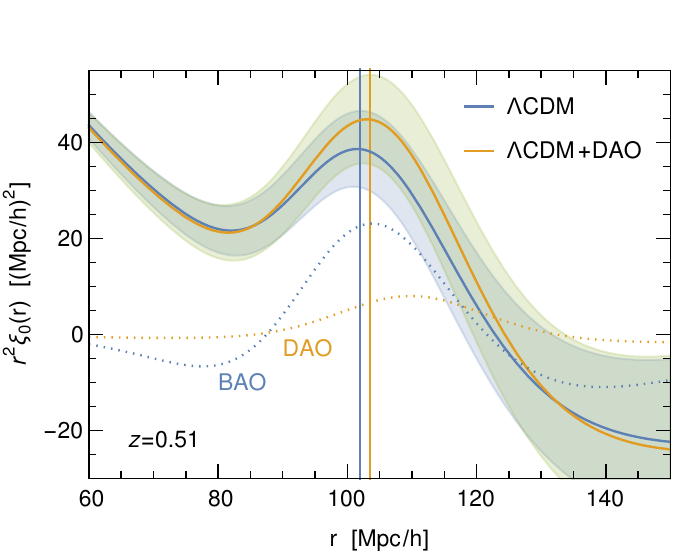}
\end{minipage}
\caption{Illustration of the monopole of the galaxy correlation function for a $\Lambda$CDM model (blue) as well as $\Lambda$CDM+DAO (orange) with parameters $r_D=1.1 \, r_B$, $\tilde A=A_D/A_B=0.25$, $\Sigma_{D,\text{Silk}}=4$ Mpc/$h$ and non-linear damping of the BAO and DAO peaks obtained from IR resummation~\cite{Blas:2016sfa,Ivanov:2018gjr}. Dotted lines are BAO and DAO peaks obtained from the corresponding oscillating component of the linear power spectrum, respectively. The offset between the blue and orange vertical lines shows the shift $\Delta\alpha_\text{iso}(z)\times r_B$ from (mis-)interpreting the combined BAO+DAO peak as BAO only according to~\eqref{eq:alphaiso}. $\Lambda$CDM parameters correspond to the Planck 2018 best fit~\cite{Planck:2018vyg}.  The shaded bands provide a rough indication for the amount of variation in peak normalization and width that could be absorbed by varying the nuisance parameters used in the fit of the theoretical template to data, assuming the prior ranges employed for the DESI analysis~\cite{DESI:2025qqy}. The redshift and galaxy bias are chosen to match those of the DESI LRG1 sample.}
\label{fig:xi0}
\end{figure}

In this work, we consider the possibility that the inferred BAO peak locations, and thus $\alpha_\parallel(z_i)$ and $\alpha_\perp(z_i)$, are biased due to the presence of a DAO peak centered around $r_{D}$, close to the BAO scale $r_{B}$, with separation that is comparable to or smaller than the total peak width. An example along these lines is illustrated in Fig.~\ref{fig:xi0}, where the monopole of the galaxy correlation function~\eqref{eq:xiell} for  parameters related to the DESI LRG1 sample (effective redshift $z=0.51$, bias $b_1=2.0$) is shown (excluding the broadband correction from the $D$-term in~\eqref{eq:Pgmodel}). The blue solid line shows a fiducial (Planck 2018) $\Lambda$CDM model, while the orange line corresponds to a model featuring a nearby DAO with $r_D=1.1 \, r_B$ and an amplitude that is $1/4$ of that for the BAO (parameter $\tilde A=A_D/A_B=0.25$, see below for details). The linear, oscillating components $P_\text{w}(k)$ for the BAO and DAO are shown by the dotted lines, and the non-linear damping scales $\Sigma_{\parallel,\perp}$ are computed using the result~\cite{Blas:2016sfa,Ivanov:2018gjr} from IR resummation for illustration. The vertical lines indicate the peak location inferred from the BAO only, and when fitting the peak position to the combined BAO+DAO peak, respectively. Their mismatch corresponds to the bias $\Delta \alpha_\text{iso}(z_i)$ in the inferred value of $\alpha_\text{iso}(z_i)$ from misinterpreting the BAO+DAO for a single BAO peak. The shaded regions around the solid lines roughly indicate the amount of variation that can be absorbed by varying the damping parameters $\Sigma_{\parallel,\perp}$ within the prior range used in the DESI analysis (Table III in~\cite{DESI:2025qqy}). They also give an approximate idea of the typical size of error bars on the monopole of the correlation function for each radial bin of the DESI DR2 data as presented in Fig.~5 in~\cite{DESI:2025zgx}, even though we caution that this depends on the galaxy sample considered.

A precise analysis of which level of DAO `admixture' could be misinterpreted by the DESI peak fitting procedure as a bias in the inferred BAO location would require a dedicated study employing the DESI pipeline (especially when including also BAO reconstruction). Here, we provide a simplified semi-analytical estimate as a basis for further exploration. For this purpose,  we consider an idealized description of the additional oscillating DAO contribution in the galaxy power spectrum described by 
\be\label{eq:DeltaPD}
\Delta P_{D}(k,\mu) = A_D e^{-\Sigma_{D}^2(z,\mu) k^2}\sin \left(k r_{D}+\theta_{D}\right) P_{\text{s}}(k,\mu)\equiv A_D W_D(k,\mu)P_{\text{s}}(k,\mu)\,,
\ee
where the DAO scale $r_D\equiv r_{d,D}$ and amplitude $A_D$ are the relevant new parameters, along with a possible phase $\theta_D$, that we however set to zero in
our numerical analysis below. $P_{\text{s}}(k,\mu)$ stands for the smooth broadband spectrum; while its precise shape is irrelevant in the following, it would be given by $(b_1+f\mu^2)^2P_\text{nw}(k)$ in linear approximation. The DAO damping is modeled by a Gaussian suppression composed of a component $\Sigma_{D,\text{Silk}}$ accounting for the possibility of attenuation of the dark acoustic sound waves in the early Universe, as well as the usual non-linear damping from bulk flows generated at late times, with redshift- and $\mu$-dependence as predicted by IR resummation~\cite{Matsubara:2007wj,Matsubara:2008wx,Carlson:2012bu,Senatore:2014via,Baldauf:2015xfa,Seo:2015eyw,Vlah:2015zda,Blas:2016sfa,Ivanov:2018gjr}\footnote{We checked that the additional contribution $\delta\Sigma^2$ (see e.g.~(7.7) in~\cite{Ivanov:2018gjr}) has no impact on our analysis, and thus neglect it.},
\be\label{eq:Sigma_decomp_dark}
\Sigma_D^2(z,\mu) = \Sigma_{D0}^2 D^2(z)(1+\mu^2f(z)(2+f(z)))+ \Sigma_{D,\text{Silk}}^2\,,
\ee
with parameters $\Sigma_{D0}$ and $\Sigma_{D,\text{Silk}}$, and redshift-dependent non-linear damping entering via the growth factor $D(z)$ and rate $f(z)=d\ln D/d\ln a$. To be able to derive an analytic expression for the apparent shift of the inferred BAO position in the presence of DAO, we also consider an analogous ansatz for the BAO contribution, and write 
\be
  P_g(k,\mu) = (1+A_BW_B(k,\mu)+A_DW_D(k,\mu))P_\text{s}(k,\mu)\,,
\ee
with BAO amplitude $A_B\simeq 0.05$ and $W_B$ analogously to~\eqref{eq:DeltaPD}, with
BAO scale $r_B$, phase $\theta_B\simeq 0$, and Gaussian damping strength
\be\label{eq:Sigma_decomp}
\Sigma^2(z,\mu) = \Sigma_0^2 D^2(z)(1+\mu^2f(z)(2+f(z)))+ \Sigma_\text{Silk}^2\,,
\ee
where the latter is the usual Silk damping scale.

To derive the bias in the inferred BAO scale we consider the Fisher matrix approximated as in~\cite{Tegmark:1997rp},
\be
{\bf F}_{ij}= \frac{\partial {\boldsymbol{\mu}}^T}{\partial \theta_i}{\bf C}^{-1} \frac{\partial {\boldsymbol{\mu}}}{\partial \theta_j}\,,
\ee
with covariance matrix ${\bf C}$. For simplicity we consider the measured parameters $\boldsymbol{\mu}$ to stand
for the power spectrum, while the parameters of interest are $\theta_i\in\{\alpha_\parallel,\alpha_\perp\}$.
Now we can estimate the difference between the best-fit value and the fiducial value for our parameters $\theta$~\cite{Bernal:2020pwq},
\be
\Delta \theta_i = {\bf F}^{-1}_{ij}b_j \,,
\ee
where we sum over repeated indices and
\be
b_j = \frac{\partial {\boldsymbol{\mu}}^T}{\partial \theta_j}{\bf C}^{-1}\Delta {\boldsymbol{\mu}}\,, 
\ee
where $\Delta {\boldsymbol{\mu}}$ is the difference between the fiducial (assumed) model and the measured true model.
For simplicity we adopt the Fisher matrix in continuous approximation from \cite{Tegmark:1997rp,Seo:2007ns}, 
\be
F_{ij} = \frac{1}{8 \pi^2} \int_{-1}^{1} d\mu \int dk k^2 \frac{\partial \ln P_g}{\partial\theta_i} \frac{\partial \ln P_g}{\partial \theta_j}V_\text{eff}(k,\mu)\,,
\ee
with effective survey volume $V_\text{eff}$.
Since BAO measurements effectively only fit the oscillating part but not the ``broadband'' smooth spectrum, which is marginalized over, one can write
\be
F_{ij} \approx \frac{1}{8 \pi^2} A_B^2\int_{-1}^{1} d\mu \int dk k^2 \frac{\partial W_B}{\partial\theta_i} \frac{\partial W_B}{\partial \theta_j}V_\text{eff}(k,\mu)\, .
\ee
Similarly, in our case, we find
\be
b_j \approx \frac{1}{8 \pi^2} A_BA_D\int_{-1}^{1} d\mu \int dk k^2 \frac{\partial W_B}{\partial\theta_j} W_D V_\text{eff}(k,\mu)\, .
\ee

\subsection{Isotropic dilatation bias}

For illustration, let us first consider a single BAO distance parameter $\alpha$
that describes an isotropic dilatation relative to the fiducial model (i.e. $\alpha\equiv\alpha_\perp=\alpha_\parallel$), before generalizing to the anisotropic case. This corresponds to a rescaling $k\to k/\alpha$ in Fourier space. Given that the fiducial model has $\alpha =1$ by definition, we  have $\partial W_B/\partial \alpha = -k \partial W_B/\partial k$. In this way we obtain
\bea\label{eq:bisodef}
b &\approx& \frac{1}{8 \pi^2} A_BA_D\int_{-1}^{1} d\mu \int dk k^2 \frac{\partial W_B}{\partial\alpha} W_D V_\text{eff}(k,\mu)\\&=& -\frac{1}{8 \pi^2} A_BA_D \int_{-1}^{1} d\mu \int dk k^3 e^{-(\Sigma^2+\Sigma_D^2)k^2}\left(r_B\cos(kr_B+\theta_B)-2\Sigma^2 k \sin(kr_B+\theta_B)\right) \sin(kr_D+\theta_D)V_{\text{eff}}(k,\mu)\, .\nn
\eea
Assuming that $r_B \gg \Sigma^2 k \sim \Sigma^2/|\Delta r|$ we can drop the second term in the bracket, and assuming $(r_B+r_D)^2\gg \Sigma^2+\Sigma_D^2$ we may keep only the parts oscillating as $k\Delta r+\Delta\theta$, with $\Delta r=r_D-r_B$ and $\Delta\theta=\theta_D-\theta_B$. Also, if $V_\text{eff}$ is assumed to vary sufficiently slowly on the relevant scales, we can evaluate it at a typical scale $k_\star$ and factor it out of the $k$-integral, obtaining
\be\label{eq:biso}
b \approx -\frac{1}{8 \pi^2} \frac{r_B}{2} A_BA_D \int_{-1}^{1} d\mu V_\text{eff}(k_*,\mu)\, {\cal I}(\Delta r,\Delta\theta,\Sigma^2+\Sigma_D^2)\,,
\ee
where
\be
 {\cal I}(\Delta r,\Delta\theta,\sigma^2) \equiv \int dk k^3 e^{-\sigma^2k^2}\sin(k\Delta r+\Delta \theta)
 \stackrel{\Delta\theta=0}{=}
 \frac{\sqrt{\pi}}{16\sigma^7}\Delta r\left[6\sigma^2-(\Delta r)^2\right]\,\exp\left[-\frac{(\Delta r)^2}{4\sigma^2}\right]\,.
\ee
In  similar approximation, the Fisher matrix can be written as
\be
F \approx  \frac{1}{8 \pi^2}\frac{r_B^2}{2} A_B^2 \int_{-1}^{1} d\mu V_\text{eff}(k_*,\mu)\, {\cal J}(2\Sigma^2), \quad {\cal J}(\sigma^2)\equiv \int dk k^4 e^{-\sigma^2 k^2} = \frac{3\sqrt{\pi}}{8\sigma^5}\, ,
\ee
which yields the shift in the inferred isotropic distance parameter\footnote{We note that the BAO damping scale could also be included in an extended Fisher analysis by enlarging the parameter vector to $\theta=(\alpha,\Sigma)$. In this case, non-vanishing mixing $F_{\alpha\Sigma}$ between the dilation and damping directions modifies the bias formula to
$
\Delta\alpha
=
\left(
F_{\alpha\alpha}
-
F_{\alpha\Sigma}F_{\Sigma\Sigma}^{-1}F_{\Sigma\alpha}
\right)^{-1}
\left(
b_\alpha
-
F_{\alpha\Sigma}F_{\Sigma\Sigma}^{-1}b_\Sigma
\right)
$.
An explicit estimate of $F_{\alpha\Sigma}$ and $b_\Sigma$, analogous to the calculation presented above for $b_\alpha\equiv b$ and $F_{\alpha\alpha}\equiv F$, shows that the additional terms are negligible if $\Sigma^2 \ll r_B^2$ and $\Sigma^2 \ll r_B\, |\Delta r|$. These conditions are well satisfied for the BAO scale, damping widths, and DAO separations considered in this work.}
\be
\boxed{
 \Delta \alpha=F^{-1}b = - \left(\frac{r_D-r_B}{r_B}\right)\times\left(\frac{A_D}{A_B}\right)\times {\cal D}(\Delta r,\Sigma,\Sigma_D) \,,
}
\ee
where the first factor is the relative shift between the DAO and BAO drag horizon, the second factor $\tilde A\equiv A_D/A_B$ the relative amplitude, and the last factor encodes the dependence on the width of the BAO and DAO peaks, $\Sigma^2$ and $\Sigma_D^2$, respectively, given by (for $\Delta\theta=0$)
\be
  {\cal D}(\Delta r,\Sigma,\Sigma_D) \equiv 
  \frac{1}{[(2\Sigma^2)^{-5/2}]_\text{av}}\,\left[
  (\Sigma^2+\Sigma_D^2)^{-5/2}\left(1-\frac{(\Delta r)^2}{6(\Sigma^2+\Sigma_D^2)}\right)\,\exp\left[-\frac{(\Delta r)^2}{4(\Sigma^2+\Sigma_D^2)}\right]\right]_\text{av}\,.
\ee
The subscript `av' stands for averaging over $\mu$, weighted with the effective survey volume, as given above.

\subsection{Anisotropic dilatation bias}

We now return to the case relevant for the DESI BAO measurement, with two independent  parameters $\alpha_\parallel$ and $\alpha_\perp$, generalizing~\eqref{eq:bisodef} to
\be\label{eq:bdef}
\left(b_\parallel\atop b_\perp\right) = \frac{1}{8 \pi^2} A_BA_D\int_{-1}^{1} d\mu \int dk k^2 \left(\partial W_B/\partial\alpha_\parallel \atop \partial W_B/\partial\alpha_\perp\right) W_D V_\text{eff}(k,\mu)\,.
\ee
From the discussion below~\eqref{eq:xiell}, $\alpha_\parallel$ and $\alpha_\perp$ enter $W_B$ via the rescalings $k_\parallel\to k_\parallel/\alpha_\parallel$ and $k_\perp\to k_\perp/\alpha_\perp$ in Fourier space.
$W_B$ depends on them via the dependence on the rescaled total wave number $k'$, and via the rescaled $\mu'$ due to the line-of-sight dependence of the Gaussian BAO damping $\Sigma^2(z,\mu')$. However, following a similar logic as after~\eqref{eq:bisodef}, the derivative acting on the latter is subdominant compared to the contribution where it acts on $\sin(k'r_B+\theta_B)$. Now, for any function of the rescaled total wavenumber $k'$, we have $\partial f(k')/\partial \alpha_\perp|_{\alpha_\perp=1} = (\partial f/\partial k)(\partial k'/\partial \alpha_\perp)|_{\alpha_\perp=1} = -(1-\mu^2)k\partial f/\partial k$ and similarly $\partial f(k')/\partial \alpha_\parallel|_{\alpha_\parallel=1} =-\mu^2k\partial f/\partial k$. From this we obtain, analogously to~\eqref{eq:biso},\footnote{We note that when disregarding the $\mu$-dependence of $V_\text{eff}$ and $\Sigma, \Sigma_D$, one would recover the isotropic case discussed before, ${\bf{F}} = F\,{\bf{\tilde F}}$ and ${\bf b} =b\,{\bf \tilde b}$ where $F$ and $b$ are the Fisher matrix and bias vector in the isotropic case above, respectively, and 
\begin{align}
{\bf{\tilde b}} ={\renewcommand{\arraystretch}{1.4}\begin{pmatrix} \frac{2}{3} \\ \frac{4}{3}  \end{pmatrix}}\,,\qquad {\bf{\tilde F}}= {
\renewcommand{\arraystretch}{1.4}
\begin{pmatrix}
\frac{2}{5} & \frac{4}{15}\\
\frac{4}{15} & \frac{16}{15}
\end{pmatrix}
}\,,\nn
\end{align}
and so ${\bf{\tilde F}}^{-1}\cdot \bf{\tilde b} = \bf 1\,$. }
\bea\label{eq:b}
\left(b_\parallel\atop b_\perp\right) 
&\approx& -\frac{1}{8 \pi^2} \frac{r_B}{2} A_BA_D \int_{-1}^{1} d\mu \,V_\text{eff}(k_*,\mu)\,\left(\mu^2\atop 1-\mu^2\right)\, {\cal I}(\Delta r,\Delta\theta,\Sigma^2(z,\mu)+\Sigma_D^2(z,\mu))\nn\,,\\
\left(\begin{array}{cc} F_{\parallel\parallel} & F_{\parallel\perp} \\ F_{\perp\parallel} & F_{\perp\perp} \end{array}\right) &\approx& \frac{1}{8 \pi^2}\frac{r_B^2}{2} A_B^2 \int_{-1}^{1} d\mu\, V_\text{eff}(k_*,\mu)\, \left(\begin{array}{cc} \mu^4 & \mu^2(1-\mu^2) \\ (1-\mu^2)\mu^2 & (1-\mu^2)^2\end{array}\right)\, {\cal J}(2\Sigma^2(z,\mu))\,.
\eea
This yields the shift in the inferred BAO distance parameters for galaxies parallel and perpendicular to the line of sight, due to the presence of DAO close to the BAO,
\be\label{eq:alphashift}
\boxed{
  \left(\Delta\alpha_\parallel(z) \atop \Delta\alpha_\perp(z) \right)
  = \left(\begin{array}{cc} F_{\parallel\parallel} & F_{\parallel\perp} \\ F_{\perp\parallel} & F_{\perp\perp} \end{array}\right)^{-1}\left(b_\parallel\atop b_\perp\right) = - \left(\frac{r_D-r_B}{r_B}\right)\times\left(\frac{A_D}{A_B}\right)\times \left({\cal D}_\parallel\atop{\cal D}_\perp\right) \,,
}
\ee
where
\be
  \left({\cal D}_\parallel\atop{\cal D}_\perp\right)\equiv \left({\cal N}\right)^{-1}\left[\left(\mu^2\atop 1-\mu^2\right)
  (\Sigma^2+\Sigma_D^2)^{-5/2}\left(1-\frac{(\Delta r)^2}{6(\Sigma^2+\Sigma_D^2)}\right)\,\exp\left[-\frac{(\Delta r)^2}{4(\Sigma^2+\Sigma_D^2)}\right]\right]_\text{av}\,,
\ee
with
\be
  {\cal N}\equiv \left[\left(\begin{array}{cc} \mu^4 & \mu^2(1-\mu^2) \\ (1-\mu^2)\mu^2 & (1-\mu^2)^2\end{array}\right)(2\Sigma^2)^{-5/2}\right]_\text{av}\,.
\ee
Note that the shifts $\Delta\alpha_\parallel(z)$ and $\Delta\alpha_\perp(z)$ pick up a dependence on redshift due to non-linear BAO damping. As mentioned above it would be interesting to apply the BAO+DAO template to the DESI pipeline including BAO reconstruction, which is beyond the scope of this work. We note that the shift predicted by~\eqref{eq:alphashift} (shown by the difference between the vertical lines in Fig.~\ref{fig:xi0}) is consistent with the shift of the position of the BAO peak in $\xi_0(r)$ as displayed in Fig.~\ref{fig:xi0}.

The shift parameter related to the angle-averaged BAO distance $D_V$ can be obtained by writing
 \be
 \alpha^{\textrm{shifted}}_i = \alpha^\textrm{base}_{i} +\Delta \alpha_i = \alpha^\textrm{base}_{i}(1+\epsilon_i)\,,
\ee
 with $\epsilon_i\equiv \Delta \alpha_i/\alpha^{\textrm{base}}_{i}$. Then
\be\label{eq:alphaiso}
\alpha^{\textrm{shifted}}_{\textrm{iso}} =  \alpha^\textrm{base}_{\textrm{iso}}(1+\epsilon_{\textrm{iso}})\quad
\text{with}\quad
\epsilon_{\textrm{iso}}= \frac{2\epsilon_{\perp} +\epsilon_{\parallel}}{3}
\ee
at linear order in the shift.

\section*{Dark acoustic oscillations and the DESI anomaly}

\begin{figure}[t]
\centering

\begin{minipage}{0.9\textwidth}
    \centering
    \includegraphics[width=0.45\textwidth]{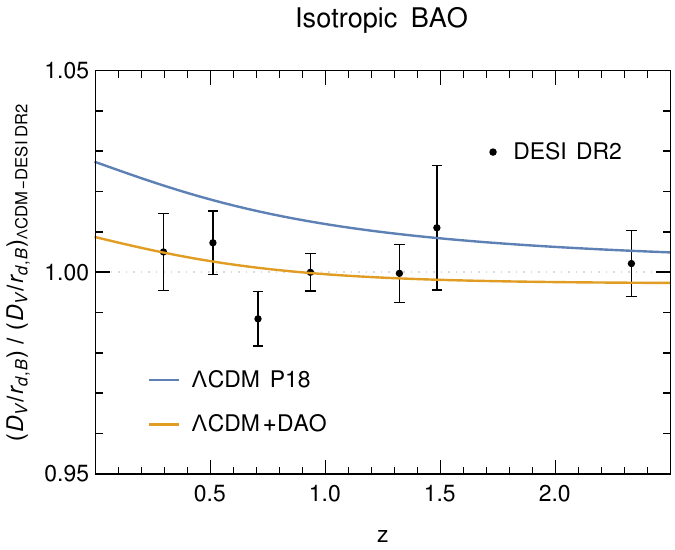}
    \quad
    \includegraphics[width=0.45\textwidth]{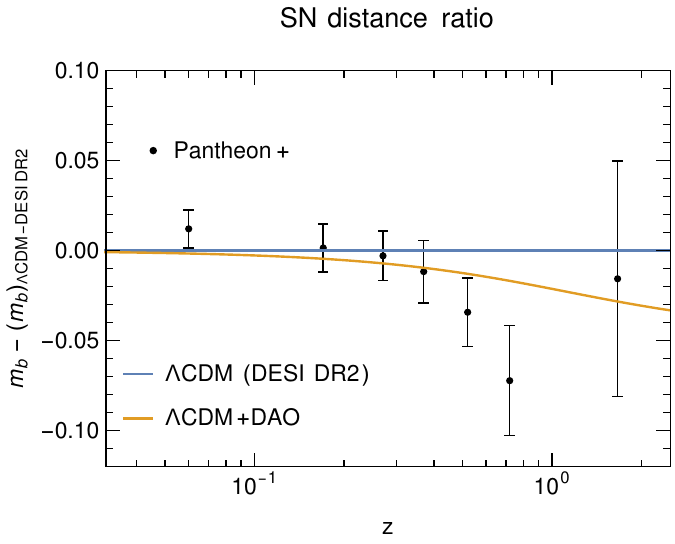}
\end{minipage}
\\[2ex]
\begin{minipage}{0.84\textwidth}
    \centering
    \includegraphics[width=\textwidth]{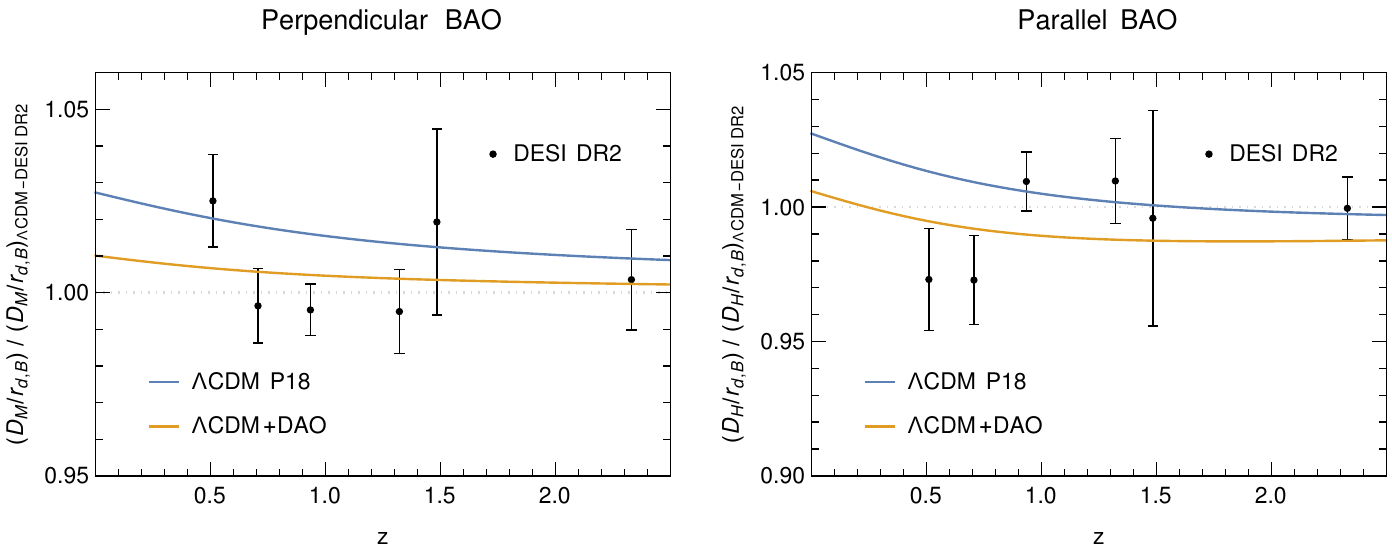}
\end{minipage}
\caption{BAO distances $D_V(z)/r_d$ (upper left), $D_M(z)/r_d$ (lower left) and $D_H(z)/r_d$ (lower right), relative to the baryon drag horizon $r_{B}$, that are inferred when (mis-)interpreting a combined BAO+DAO peak as the BAO (orange lines), compared to DESI DR2 measurements~\cite{DESI:2025zgx} as well as the prediction from a $\Lambda$CDM model with parameters matching Planck 2018~\cite{Planck:2018vyg} (blue lines). For the DAO we assume the same parameters as in Fig.~\ref{fig:xi0} for illustration, $r_D=1.1 \, r_B$, $\tilde A=A_D/A_B=0.25$, $\Sigma_{D,\text{Silk}}=4$ Mpc$/h$. The upper right panel shows SN data by Pantheon+ compared to the predictions of the two models. Notably, including the DAO in the fit to DESI DR2 BAO data allows for cosmological parameters close to those preferred by Planck (in particular for $\Omega_m$), restoring consistency with SN data as well in this way. All quantities are normalized to a standard $\Lambda$CDM model with best-fit parameters from DESI DR2.}
\label{fig:DAO_fit}
\end{figure}

We investigate to which extent the DAO effect discussed before can explain the DESI anomaly. 
As we have seen above, if the BAO measurement is slightly biased by the presence of DAO, this can be modeled in terms of a linear $z$-dependent shift $\alpha|_\mathrm{measured} \to  \alpha|_\mathrm{measured} - \Delta \alpha(z) $. Explicitly, if we define the residual vector as 
\begin{align}\label{eq:residual}
r_l = \left[\alpha|_\mathrm{theory} - (\alpha|_\mathrm{measured}-\Delta \alpha)\right]_l\,,
\end{align}
where $l$ labels different BAO measurements and $\Delta \alpha(z) \in\{\Delta \alpha_\mathrm{\perp},\,\Delta \alpha_\mathrm{\parallel}\}$ is defined in \eqref{eq:alphashift}, the total $\chi^2$ is
\begin{align}\label{eq:chi_sq}
\chi^2 = \sum_{l,l^\prime}r_l\left(C^{-1}\right)_{ll^\prime}r_{l^\prime}\,,
\end{align}
where $C$ is the covariance matrix. Thus, due to \eqref{eq:residual}, in the presence of generic DAO, the naive theory prediction valid in their absence should be replaced, $\alpha|_\mathrm{theory} \to  \alpha|_\mathrm{theory} + \Delta \alpha(z) $, when comparing with \textit{regular} BAO data. We demonstrate how this affects the fit to BAO and SN data in Fig.~\ref{fig:DAO_fit}, where we use the Planck 2018 best fit $\Lambda$CDM cosmology as reference~\cite{Planck:2018vyg}. While the blue curve corresponds to the best fit $\Lambda$CDM cosmology in the absence of the DAO feature when fitted to  Planck 2018 data~\cite{Planck:2018vyg}, the orange curve includes the DAO feature, resulting in a visibly improved fit to DESI DR2 data for isotropic, as well as perpendicular and parallel BAO distances. Notably, the same model additionally restores consistency with SN data (upper right panel in Fig.~\ref{fig:DAO_fit}). This is because the apparent shift in BAO position due to the DAO admixture allows to fit DESI DR2 data also for larger $\Omega_m$ values, with the latter being preferred by SN.

To obtain a more quantitative understanding of how strongly the DAO feature can improve the fit to DESI, we also performed an MCMC analysis, including Planck 2018 data (low-$\ell$ TT+EE, high-$\ell$ TT+TE+EE, and lensing)~\cite{Planck:2018vyg,Planck:2018lbu}, Release 2 (DR2) DESI BAO data~\cite{DESI:2025zgx}, and Pantheon+ SN data~\cite{Brout:2022vxf}. We implemented the DAO feature in a modified DESI likelihood for \texttt{Cobaya}~\cite{Torrado:2020dgo}, using \eqref{eq:chi_sq} together with \eqref{eq:alphashift}, where we neglected the $\mu$-dependence of $V_\mathrm{eff}$.\footnote{The likelihood is publicly available at \href{https://github.com/NEDE-Cosmo/DAO_bias_DESI_DR2}{github.com/NEDE-Cosmo/DAO\_bias\_DESI\_DR2}.} As can be seen from \eqref{eq:Sigma_decomp} and \eqref{eq:Sigma_decomp_dark}, this implementation introduces the following set of additional parameters: The difference in drag horizon between DAO and BAO $\Delta r/r_B = (r_D - r_B)/r_B$, the relative DAO amplitude $\tilde A=A_D/A_B$, as well as the three damping parameters $h \Sigma_0$, $h \Sigma_{D0}$, and $h\Sigma_{D,\text{Silk}}$. On the other hand, the visible sector Silk damping scale $h\Sigma_\text{Silk}$ and the baryon drag horizon $r_B$ (alongside other background quantities such as the growth factor $D(z)$ and rate $f(z)$ as well as distance scales $D_{H/M/V}$) are extracted from the cosmological simulation of the $\Lambda$CDM model, performed with  \texttt{CLASS}~\cite{Blas:2011rf}. In our analysis, we vary $h \Sigma_{D,\text{Silk}}$ in the range $[0,18\,\mathrm{Mpc}]$ while we fix $h \Sigma_0 = 5\, \mathrm{Mpc}$, $h \Sigma_{D0}=5\, \mathrm{Mpc}$ for faster convergence, although we checked that allowing these two parameters to vary within the range $h\Sigma_{0},h \Sigma_{D0} \simeq 3\text{--}6\,\mathrm{Mpc}$ will not affect the outcome significantly.\footnote{This range is compatible with the typical values of $h\Sigma_{0}$ inferred from the literature, for instance from Tab.~III of Ref.~\cite{DESI:2025qqy} or Fig.~3 of Ref.~\cite{Ivanov:2018gjr}. Performing a test analysis with these extended priors, we obtain for the negative branch $\Delta r/r_B=-0.44^{+0.12}_{-0.15} $, which is fully compatible with the value $\Delta r/r_B=-0.42^{+0.08}_{-0.11}$ reported here.  We stress, however, that the theoretical uncertainty in $h\Sigma_{D0}$ is larger than for $h\Sigma_{0}$, and that a dedicated study of the nonlinear damping in the dark sector fluid would be valuable.} The relative amplitude $\tilde A$ is varied in the range $[0,1]$. As explained in the analytic analysis, the relative difference between DAO and BAO drag horizon $\Delta r/r_B$ features a bi-modality and is therefore split into two regimes, where the negative branch is explored in the range $[-0.6,0]$ and the positive branch in $[0,0.4]$. For the remaining six $\Lambda$CDM parameters, we choose priors compatible with standard choices in the literature~\cite{Planck:2018vyg}.
Beyond the Bayesian analysis, we  also compute a profile likelihood curve along the $\tilde A$ direction, using the \texttt{Prospect} code~\cite{Holm:2023uwa}. This approach evaluates the likelihood of the data given fixed choices of the model parameters and has the benefit of being less sensitive to prior-volume effects.

\begin{table*}[t]
\renewcommand{\arraystretch}{1.7}
	\centering
    \fontsize{8}{9.5}\selectfont
		\begin{tabular}{ | l || c | c | c | c |c|}
			\hline
			&$\Delta r/r_B$&$\tilde A$&$h\,   r_{B}\, \mathrm{[Mpc]}$&$\Omega_m$ &$\Delta \chi^2$\\ \hline
			
			$\Lambda$CDM  &--& --   & $100.79\pm 0.48$&\makecell{$0.3021\pm 0.0036$\\ ($0.3007$)}&$0$
			\\ \hline 
			\makecell{+ DAO (pos.~branch)} &\makecell{$0.17_{-0.11}^{+0.10}$\\ ($0.167$)}& \makecell{  $(\ast)$\\($0.234$)}& $100.04^{+0.70}_{-0.62}$ & \makecell{$0.3078^{+0.0047}_{-0.0055}$\\ ($0.3099$)}&$ -4.3$
			\\ \hline 

   			+ DAO (neg.~branch) &\makecell{$-0.42_{-0.11}^{+0.08}$\\($-0.436$)}&\makecell{  $(\ast)$\\($0.960$)} &$100.34^{+0.68}_{-0.57}$&\makecell{ $0.3055^{+0.0043}_{-0.0053}$\\ ($0.3070$)} & $-4.7$ \\ \hline   			+\,($w_0$,$w_a$) &--&--  & $99.64\pm 0.86$ &\makecell{ $0.3105\pm 0.0056$\\ ($0.3072$)} & $-8.7$ \\			\hline
			
		\end{tabular}
	\caption{Main results of combined analysis with  CMB (Planck 2018), SN (Pantheon+), and BAO (DESI DR2) data. For a more comprehensive summary see Tab.~\ref{tab:MCMC}.  If not stated otherwise, we report the $68 \%$ CL and best fit values in parentheses.\\ $(\ast)${ Unconstrained by Bayesian analysis within $95 \%$ CL. We refer in these cases to the profile likelihood analysis in Fig.~\ref{fig:DAO_profile}.  }}
	\label{tab:results}
\end{table*}

\begin{figure}[t!]
    \centering
    \includegraphics[width=0.95\linewidth]{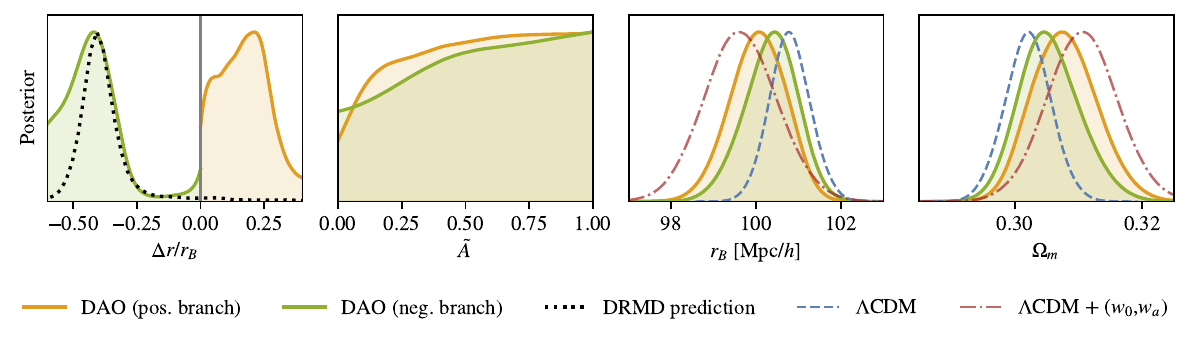}
    \vspace{-0.1cm}
    \caption{1D posterior distributions of the DAO  parameters $\Delta r/r_B=(r_{D}-r_{B})/r_B$ and $\tilde A=A_D/A_B$, as well as $ r_{B}$, and $\Omega_m$ from a combined analysis with Planck 2018, DESI DR2, and Pantheon+. The data fit shows a preference for a non-vanishing $\tilde A$ corresponding to DAO with an approximately $10- 20\%$ larger (`positive branch' in orange) or $ 40\%$ smaller (`negative branch' in green) dark sound horizon $r_{d,D}\equiv r_{D}$. The black dotted line depicts the DAO prediction in the DRMD model obtained in~\cite{Garny:2025kqj} (with an additional prior on the absolute SN magnitude taken from~\cite{Riess:2021jrx}). Beyond improving the fit to DESI data, the DAO feature leads to larger values of $\Omega_m$, also improving the fit to SN data.}
    \label{fig:DAO_posterior}
\end{figure}

A summary of our statistical results, including the two most relevant DAO parameters, is presented in Tab.~\ref{tab:results} and Fig.~\ref{fig:DAO_posterior} (a more complete compilation of the MCMC results is provided in Tab.~\ref{tab:MCMC} in the Appendix). For both the positive and negative branches, we find an improvement in the fit to the data, with $\Delta\chi^2 = -4.3$ and $-4.7$, respectively. This is consistent with the behavior shown in Fig.~\ref{fig:DAO_fit}, where the orange curves (corresponding to a positive-branch DAO) provide a visibly better fit to the BAO measurements than the blue $\Lambda$CDM curve. We also observe an improved fit to the supernova data (upper-right panel), which we attribute to a positive shift in $\Omega_m$ compared to the $\Lambda$CDM analysis (see also the green dashed contour in the rightmost panel in Fig.~\ref{fig:DAO_posterior}). While this is a weaker improvement than that obtained in the $(w_0,w_a)$ extension (for which $\Delta \chi^2=-8.7$), we stress that the presence of DAO can in principle be founded in concrete particle physics. At this stage, we also refrain from performing a model comparison beyond this simple $\chi^2 $ statistics. The reason is that the DAO should arise in a more complete dark sector model, such as DRMD, where ultimately $\tilde A$, $\Delta r/r_B$, and $\Sigma_{D,\mathrm{Silk}}$ are fixed in terms of the underlying model parameters.\footnote{In this context, we refer the reader to the dedicated follow-up analysis in~\cite{Garny:2026ish} which studies the imprint of DAO on the CMB within the DRMD model. In particular, it is found that Planck CMB data are compatible with rather large values of $\tilde A \lesssim 0.8 $. This implies that the typical range considered in this work, $0.1 \lesssim \tilde A \lesssim 0.5$, does not require extreme deviations from $\Lambda$CDM at early times.}

\begin{figure}[t!]
    \centering
    \parbox{0.49\linewidth}{
        \centering
        \includegraphics[width=\linewidth]{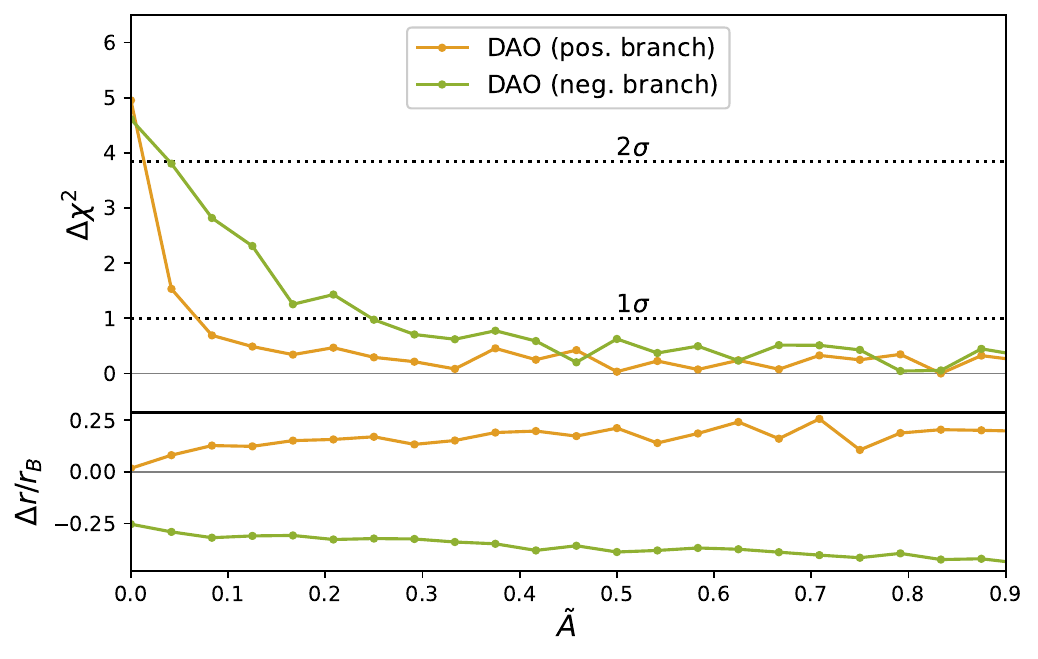}
    }
    \hfill
    \parbox{0.49\linewidth}{
        \centering
        \includegraphics[width=\linewidth]{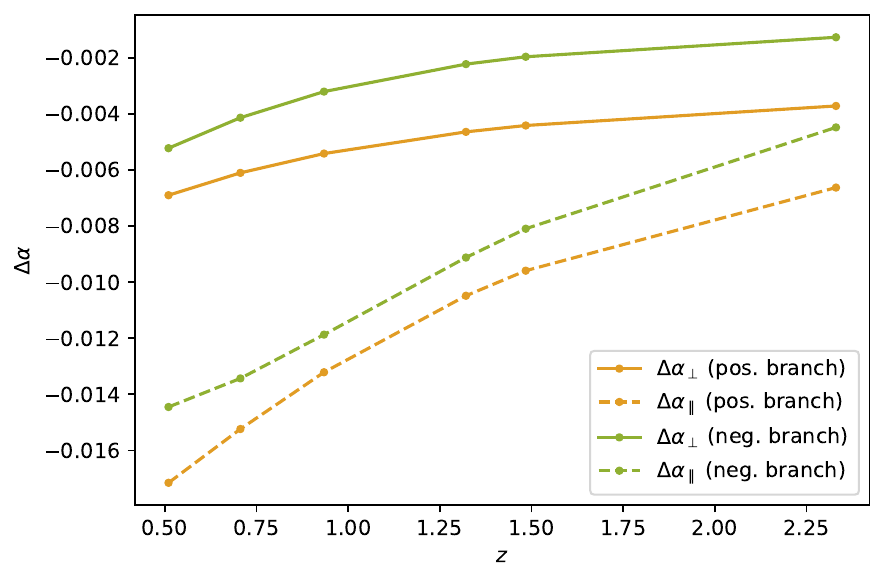}
    }
    \vspace{-0.15cm}
    
    \caption{Comparison between the positive (orange) and negative (green) DAO branches, which correspond to positive and negative values of $\Delta r / r_B$, respectively. \underline{Left:} Profile likelihood curves for Planck 2018, DESI DR2, and Pantheon+ data within $\Lambda$CDM, allowing for the presence of DAO with relative amplitude $\tilde A$, which biases the inference of the baryon drag horizon $r_B$. The second row depicts the relative difference between dark and visible sector drag horizon scales $\Delta r/r_B=(r_{D}-r_{B})/r_B$. For the positive branch a small relative displacement between DAO and BAO peaks is enough to improve the fit significantly. For the negative branch, the DAO tail is responsible for shifting the apparent BAO peak and thus a larger displacement and amplitude is necessary to achieve a similar effect. \underline{Right:} Strength of the DAO bias effect as a function of redshift for the best fit points from the left panel at $\tilde A = 0.5$. Overall the bias effect is more pronounced for parallel BAO (dashed) and small redshift.}
    \label{fig:DAO_profile}
\end{figure}

The DAO amplitude is parametrized by $\tilde A$.
From the posterior shown in Fig.~\ref{fig:DAO_posterior}, it appears only weakly constrained, with a mild preference for values of order unity. This behavior, however, is largely driven by the Bayesian analysis not efficiently probing the region near $\tilde A=0$.
To clarify this point, we performed a profile-likelihood analysis, the results of which are shown in Fig.~\ref{fig:DAO_profile}.
Here, the point $\tilde A = 0$, corresponding to the absence of a DAO feature, lies outside the reconstructed $2\, \sigma$ region for both the positive (orange) and negative (green) branches. 
The corresponding preferred regions are $\tilde A \gtrsim 0.1$ for the positive branch and $\tilde A \gtrsim 0.3$ for the negative branch. The larger preferred amplitude in the negative branch can be understood physically: since this solution relies primarily on the large-scale tail of the DAO peak to change the BAO peak position, a larger amplitude is required to achieve a comparable improvement in the fit relative to the positive branch for which both peaks overlap.

We further observe, in both the one-dimensional posterior and the profile likelihood, that the likelihood becomes very flat for $\tilde A > 0.4$ in both branches.
This behavior can be traced to an approximate degeneracy between $\tilde A$ and  $\Sigma_{D,\mathrm{Silk}}$, whereby an increased DAO amplitude can be compensated by stronger damping.
We expect this degeneracy to be at least partially lifted in a more complete (nonlinear) analysis, and therefore do not assign strong physical significance to the large-amplitude regime. In fact, such large values of $\tilde A$ may already be in tension with full-shape constraints. Therefore, the physically interesting region where the DAO bias provides an interesting scenario for resolving the DESI anomaly is for $0.1 \lesssim \tilde A \lesssim 0.5$, which corresponds to a percent-level absolute amplitude of the DAO feature in the linear spectrum, $0.005\lesssim A_D\lesssim 0.025$. 

In the right panel of Fig.~\ref{fig:DAO_profile}, we show the redshift dependence of the parallel (dashed) and perpendicular (solid) bias effects, $\Delta \alpha_\parallel$ and $\Delta \alpha_\perp$, for the maximum-likelihood cosmology with $\tilde A = 0.5$, corresponding to one of the points in the profile likelihood. The effect decreases substantially in magnitude with increasing redshift and exhibits a pronounced anisotropy at low redshift. This characteristic redshift dependence is what improves the fit to data and renders DESI data compatible with a larger value of $\Omega_m$.

We find that the relative difference of the dark and visible sector drag horizons $\Delta r/r_B$ is relatively tightly constrained.
For the positive branch (orange posterior), we have
$\Delta r / r_B = 0.17^{+0.10}_{-0.11}$, indicating a $10-20\%$ displacement between the BAO and DAO peaks.
As discussed above and illustrated in Fig.~\ref{fig:xi0}, this corresponds to the DAO peak partially overlapping the BAO feature, resulting in a modest net shift toward larger scales.
For the negative branch (green posterior), on the other hand, we obtain $\Delta r / r_B = -0.42^{+0.08}_{-0.11}$, which is a significantly stronger and opposite shift. In this case, both peaks are clearly separated, with the DAO peak located at smaller scales than its BAO counterpart; nevertheless, as mentioned above, the large-scale tail of the DAO peak suppresses power in such a way that the apparent BAO peak is shifted toward larger scales. In both scenarios, the true baryon drag horizon $r_{B}$, which accounts for the DAO-induced bias, is smaller than that predicted in a pure $\Lambda$CDM cosmology
(see also the comparison between the orange/green solid and the blue dashed curve in Fig.~\ref{fig:DAO_posterior}).

In the case of the negative branch, we can compare our inferred value of $\Delta r/r_B$ with the prediction of the DRMD model~\cite{Garny:2025kqj}. In this scenario, dark matter and dark radiation decouple shortly before recombination, a feature that allows to resolve the Hubble tension with the standard data sets used in this work. Microscopically, this late decoupling is explained by a small mass splitting $\Delta m$ between charged and neutral dark matter particles, which leads to an additional temperature dependence of the interaction rate that scales as $\exp{\left(-\Delta m/T_d\right)}$ and suppresses the drag rate between both fluids for dark sector temperatures $T_d < \Delta m$. Eventually both fluids decouple at redshift $z_\mathrm{dec}$, where Planck data prefer values around $z_\mathrm{dec} \sim 2000$.  This late decoupling affects the gravitational potential on small scales. Crucially, it permits larger values of $\omega_b$ and $\omega_\mathrm{cdm}$, which are typically required by early-time solutions to the Hubble tension, without generating excess small-scale power. A key phenomenological signature of this mechanism is the presence of DAO with a characteristic scale set by the redshift of decoupling, $z_\mathrm{dec}$, given by
\begin{align}\label{eq:r_D}
r_D = \int_{z_\mathrm{dec}}^\infty \mathrm{d}z\, \frac{c_{s,\mathrm{eff}}}{H(z)}\,,
\end{align}
where $c^{-2}_{s,\mathrm{eff}} = 3(1+R)$ denotes the sound speed of the coupled dark matter–dark radiation fluid and $R = 3\rho_\mathrm{idm}/(4\rho_\mathrm{DR})$ depends on the energy density of interacting dark matter and dark radiation.

To infer the posterior distribution of $\Delta r/r_B$ within the DRMD model, we evaluated Eq.~\eqref{eq:r_D} in a post-processing of the  MCMC chains from~\cite{Garny:2025kqj}. These chains were obtained in a combined analysis of the same datasets employed here (Planck 2018, DESI DR2, Pantheon+), supplemented by the SH$_0$ES prior on the absolute SN magnitude $M$. The inclusion of the SH$_0$ES measurement is motivated by our interest in the region of parameter space where the model provides a resolution to the Hubble tension. Crucially, this dataset combination has been shown to be statistically consistent within the DRMD model. The resulting posterior is shown as the black dotted line in the left panel of Fig.~\ref{fig:DAO_posterior} and corresponds to $\Delta r/r_B \in [-0.48,-0.32] $ (68\% CL).

As a result, we find that the DAO-based constraint on $\Delta r/r_B$ (derived assuming a $\Lambda$CDM cosmology) appears consistent with the prediction obtained from the DRMD model, which modifies $\Lambda$CDM only at early times. We note that the DAO bias analysis performed here has limited sensitivity to early-Universe physics (with most of its dependence entering through $\Omega_m$ and $h\,r_B$), and thus should be broadly applicable to early-time models such as DRMD. We interpret this agreement as an encouraging first piece of evidence in support of the DAO hypothesis proposed in this work. A more complete analysis within the DRMD model would be required to more firmly establish this connection. Intriguingly, in such a scenario the requirement of solving the Hubble tension is expected to impose a lower bound on $\tilde A$, making the DAO feature a falsifiable prediction, potentially testable with future full-shape data from DESI or Euclid.

\begin{figure}[t!]
    \centering
    \parbox{0.49\linewidth}{
        \centering
        {$\Lambda$CDM}\\[-0.5pt]
        \includegraphics[width=\linewidth]{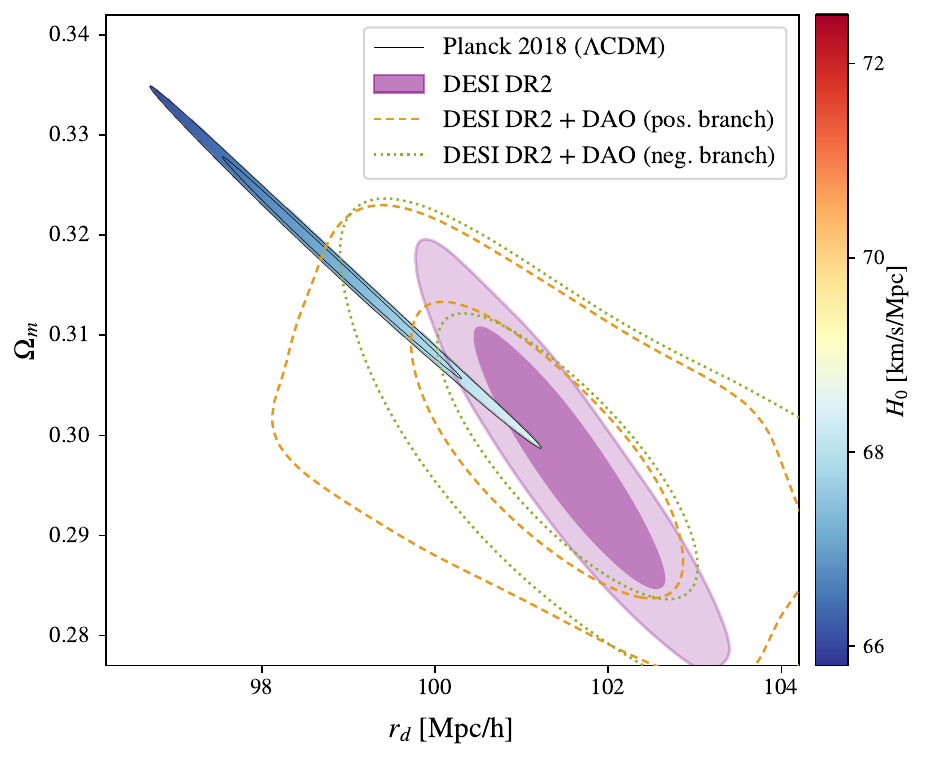}
    }
    \hfill
    \parbox{0.49\linewidth}{
        \centering
        {DRMD}\\[-0.5pt]
        \includegraphics[width=\linewidth]{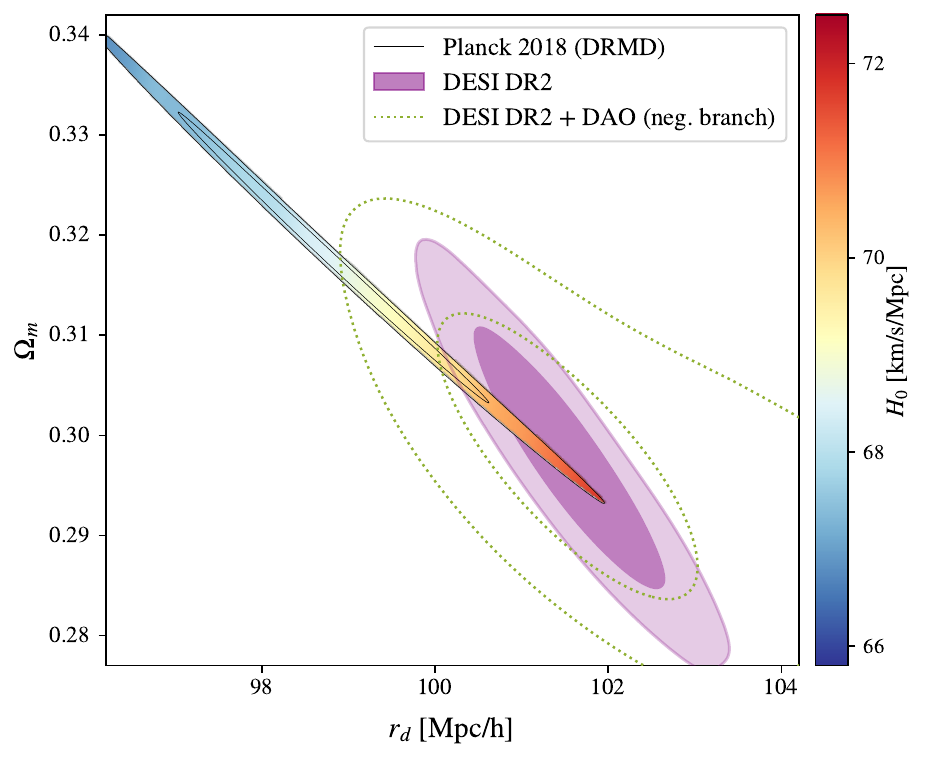}
    }
    \vspace{-0.15cm}
    \caption{
Constraints on $\Omega_m$ and the baryon drag horizon $r_B$.
Contours denote the $68\%$ and $95\%$ confidence regions; for the DESI analysis, we adopt a BBN prior on $\omega_b$ following~\cite{DESI:2025zgx}.
\underline{Left:}
Within $\Lambda$CDM, Planck 2018 data (color-coded contours) exhibit a mild tension with DESI DR2 measurements (purple contours) at the $2.3\,\sigma$ level.
This tension is alleviated when the DAO feature is included (orange dashed and green dotted contours), which corrects the misidentification of $r_{B}$ and shifts the DESI constraints toward larger values of $\Omega_m$.
\underline{Right:}
Within the DRMD model, also the CMB contours extend further into the DESI-preferred region, particularly at larger values of $H_0$. 
}
    \label{fig:DESI_CMB_tension}
\end{figure}

Finally, we turn to the so-called CMB--DESI tension, which refers to a mild
$2.3\,\sigma$ discrepancy between Planck 2018 (temperature, polarization, and
lensing) data and DESI DR2 measurements in the
$\Omega_m$ vs $h\,r_{B}$ plane within the $\Lambda$CDM model~\cite{DESI:2025zgx}. This is illustrated by the color-coded (Planck) and purple (DESI) contours in the left panel of Fig.~\ref{fig:DESI_CMB_tension}. Specifically, DESI favors lower values of $\Omega_m$ and larger values of $r_{B}$ than those preferred by Planck.
Including the DAO feature leaves the CMB predictions unchanged, while modifying
the BAO parameter inference.
In particular, the inferred value of $\Omega_m$ increases and that of
$r_{B}$ decreases, thereby reducing the tension.
For both the positive (orange dashed contours) and negative branch (green dotted contours), this shift leads to an overlap of the $68\%$ confidence regions, hence resolving the (mild) tension.

This interpretation assumes a $\Lambda$CDM description of the early
Universe. However, the existence of DAO itself would point to new pre-recombination physics,
such as Atomic Dark Matter or DRMD. While a complete and self-consistent analysis of DAO within these models is beyond the scope of this work, we provide a first consistency check in the right
panel of Fig.~\ref{fig:DESI_CMB_tension}, where we show DRMD constraints (color-coded) from a Planck-only CMB analysis alongside the DESI contours. The DESI constraints are obtained as before by assuming $\Lambda$CDM, which nonetheless provides a good approximation to DRMD at late times. Notably, for sufficiently large values of the Hubble parameter,
$H_0 \gtrsim 70\,\mathrm{km\,s^{-1}\,Mpc^{-1}}$,
the CMB-inferred constraints extend into the region preferred by DESI.
Beyond the DAO feature, this observation provides an additional connection between the DESI discrepancy and early-Universe solutions to the Hubble tension.
We stress that this comparison should be viewed as preliminary.
In particular, the present analysis does not yet use the DRMD simulation to inform
the DAO parameters $\tilde A$, $r_{d,D}$, and $\Sigma_{D,\mathrm{Silk}}$, nor does it incorporate full-shape constraints on additional features in the matter power spectrum.

\section*{Summary and discussion }

The DESI DR2 data release shows anomalies when compared to CMB and SN data within the $\Lambda$CDM cosmological model, which has been interpreted by the collaboration itself as evidence for evolving DE. Here we offer an alternative interpretation in terms of early-Universe physics, connecting the DESI anomaly to a possible solution to the Hubble tension within either the DRMD or Atomic Dark Matter models, which predict the existence of a DAO feature close in scale to the BAO~\cite{Buen-Abad:2024tlb,Garny:2025kqj,Buen-Abad:2025bgd}. 

We find that the existence of a DAO feature can bias the values of the inferred BAO distance parameters $\alpha_\perp(z)\propto D_M(z)/r_d$ and $\alpha_\parallel(z)\propto D_H(z)/r_d$ if the measured peak in the galaxy correlation function is misinterpreted as being due to BAO only. In this case one would actually expect that the distances $\alpha_i(z)$ measured via this technique differ from the ones expected from the CMB anisotropies, even if the evolution of the Universe after recombination precisely matches with $\Lambda$CDM. This is because the DAO imprints itself in the DM distribution and thus the galaxy correlations, while having no direct impact on the CMB acoustic peak. 

We derive an approximate analytical method for taking into account the impact of a DAO feature close in scale to the BAO on the distance measurement, and derive explicit expressions for the apparent shifts $\Delta\alpha_i(z)$. They depend on the relative offset $\Delta r/r_B=(r_D-r_B)/r_B$ of DAO and BAO peak scales as well as the amplitude ratio $A_D/A_B$, and become exponentially suppressed if $\Delta r$ exceeds the total peak width. Non-linear broadening leads to a redshift-dependence of $\Delta\alpha_i(z)$. Using this approach, we find that the presence of a DAO can indeed reconcile DESI DR2 with Planck 2018 as well as Pantheon+ SN data. We consider two scenarios, one with a DAO on slightly larger scale than the BAO (termed `positive branch'), and one for which the DAO is on smaller scales but has a characteristic tail overlapping with the BAO (`negative branch'). 

With these data sets the fit improves by $\Delta \chi^2 \simeq -5$  relative to $\Lambda$CDM for the positive and negative branch scenarios. This is comparable to the case when considering dynamical DE with phantom crossing (giving $\Delta \chi^2 \simeq-9$ for the same data set combination). The amplitude of the DAO is at percent level, and can be searched for with future full-shape galaxy survey data from DESI or Euclid. For the positive branch the two peaks significantly overlap, while the negative branch features a shallow DAO peak on scales $\sim 50$Mpc$/h$, just outside the range considered by DESI. Our results motivate a dedicated analysis of this scenario based on adapting the full DESI pipeline, including BAO reconstruction.
Intriguingly, the properties of the DAO required for the negative branch scenario closely match a previous prediction within the DRMD model for resolving the Hubble tension \cite{Garny:2025kqj}. Therefore, the DESI anomaly could be just another manifestation of the same physics explaining the discrepancy of CMB/BAO versus SH$_0$ES data.

\subsection*{Acknowledgements}

We thank Benjamin Bose, Shi-Fan Chen, Mikhail Ivanov, Henrique Rubira, and Benjamin Wallisch for useful conversations.
FN was supported by VR Starting Grant 2022-03160 of the Swedish Research Council.  MG acknowledges support
by the DFG Collaborative Research Institution Neutrinos and Dark Matter in Astro- and
Particle Physics (SFB 1258) and the Excellence Cluster ORIGINS - EXC-2094 - 390783311. MSS acknowledges Nordita for their kind hospitality through the Nordita corresponding fellow program.

\newpage
\begin{appendix}

\section*{Summary of MCMC results}

The detailed results of our MCMC analysis are summarized in Tab.~\ref{tab:MCMC}.

\begin{table}[h!]
\centering
\fontsize{8}{9.5}\selectfont
\setlength{\tabcolsep}{7pt} 
\renewcommand{\arraystretch}{1.2} 
\begin{tabular}{l|c|c|c|c}  
 &\textbf{$\Lambda$CDM}&\textbf{+DAO (pos. branch)}&\textbf{+ DAO (neg. branch)}& \textbf{+($w_0,w_a$)}\\
\hline\hline
\rule{0pt}{4ex}\textbf{Parameter}& \makecell{$68 \%$ limits  \\(best fit)}  & \makecell{$68 \%$ limits  \\(best fit)}  & \makecell{$68 \%$ limits  \\(best fit)} & \makecell{$68 \%$ limits  \\(best fit)}\\ 
\hline
\rule{0pt}{4ex}\makecell{$\omega_b$} & \makecell{$0.02251\pm 0.00013$\\$(0.02251)$} & \makecell{$0.02245\pm 0.00013$\\$(0.02242)$} & \makecell{$0.02247\pm 0.00013$\\$(0.02245)$} & \makecell{$0.02243\pm 0.00013$\\$(0.02244)$} \\
\hline
\rule{0pt}{4ex}\makecell{$\omega_{\mathrm{cdm}}$} & \makecell{$0.11786\pm 0.00063$\\$(0.11787)$} & \makecell{$0.11880^{+0.00081}_{-0.00090}$\\$(0.11941)$} & \makecell{$0.11842^{+0.00074}_{-0.00088}$\\$(0.11892)$} & \makecell{$0.11899\pm 0.00088$\\$(0.11874)$} \\
\hline
\rule{0pt}{4ex}\makecell{$H_0\,\mathrm{[km/sec/Mpc]}$} & \makecell{$68.33\pm 0.29$\\$(68.32)$} & \makecell{$67.90^{+0.41}_{-0.36}$\\$(67.65)$} & \makecell{$68.07^{+0.40}_{-0.33}$\\$(67.86)$} & \makecell{$67.64\pm 0.59$\\$(67.80)$} \\
\hline
\rule{0pt}{4ex}\makecell{$\ln(10^{10} A_s)$} & \makecell{$3.052\pm 0.015$\\$(3.055)$} & \makecell{$3.048^{+0.013}_{-0.015}$\\$(3.045)$} & \makecell{$3.049\pm 0.015$\\$(3.048)$} & \makecell{$3.044\pm 0.014$\\$(3.045)$} \\
\hline
\rule{0pt}{4ex}\makecell{$n_\mathrm{s}$} & \makecell{$0.9698\pm 0.0033$\\$(0.9715)$} & \makecell{$0.9675\pm 0.0036$\\$(0.9670)$} & \makecell{$0.9684\pm 0.0035$\\$(0.9679)$} & \makecell{$0.9670\pm 0.0036$\\$(0.9679)$} \\
\hline
\rule{0pt}{4ex}\makecell{$\tau_{\mathrm{reio}}$} & \makecell{$0.0600^{+0.0067}_{-0.0076}$\\$(0.0607)$} & \makecell{$0.0573^{+0.0066}_{-0.0074}$\\$(0.0548)$} & \makecell{$0.0583\pm 0.0073$\\$(0.0573)$} & \makecell{$0.0554\pm 0.0072$\\$(0.0557)$} \\
\hline
\rule{0pt}{4ex}\makecell{$\Delta r/r_B$} & -- & \makecell{$0.170^{+0.096}_{-0.11}\,(68\%\,\mathrm{CL})$\\$< 0.330\,(95\%\,\mathrm{CL})$\\$(0.167)$} & \makecell{$-0.420^{+0.084}_{-0.11}$\\$(-0.436)$} & -- \\
\hline
\rule{0pt}{4ex}\makecell{$\tilde A$} & -- & \makecell{$(*)$\\$(0.234)$} & \makecell{$(*)$\\$(0.960)$} & -- \\
\hline
\rule{0pt}{4ex}\makecell{$h \Sigma_{D,\mathrm{Silk}} \,\mathrm{[Mpc]}$} & -- & \makecell{$11.3^{+4.7}_{-3.0}\,(68\%\,\mathrm{CL})$\\$> 4.19\,(95\%\,\mathrm{CL})$\\$(6.3)$} & \makecell{$7.5^{+2.3}_{-7.1}$\\$(8.1)$} & -- \\
\hline
\rule{0pt}{4ex}\makecell{$h\, r_{B}\, \mathrm{[Mpc]}$} & \makecell{$100.79\pm 0.48$} & \makecell{$100.04^{+0.70}_{-0.62}$} & \makecell{$100.34^{+0.68}_{-0.57}$} & \makecell{$99.64\pm 0.86$} \\
\hline
\rule{0pt}{4ex}\makecell{$h\, r_{D}\, \mathrm{[Mpc]}$} & -- & \makecell{$117.1^{+8.5}_{-12}$} & \makecell{$58^{+8}_{-10}$} & -- \\
\hline
\rule{0pt}{4ex}\makecell{$\Omega_m$} & \makecell{$0.3021\pm 0.0036$\\$(0.3007)$} & \makecell{$0.3078^{+0.0047}_{-0.0055}$\\$(0.3099)$} & \makecell{$0.3055^{+0.0043}_{-0.0053}$\\$(0.3070)$} & \makecell{$0.3105\pm 0.0056$\\$(0.3072)$}\\
  \hline
  \rule{0pt}{4ex}\makecell{$\mathbf{\chi^2}$ \textbf{per} \\ \textbf{experiment}}&  &  &  \\ 
\hline\hline
\rule{0pt}{3ex}Planck high-$\ell$ & 2345.1 & 2343.3 & 2342.9 & 2342.9 \\
\rule{0pt}{3ex}Planck low-$\ell$ EE & 397.5 & 396.2 & 396.6 & 396.2 \\
\rule{0pt}{3ex}Planck low-$\ell$ TT & 22.3 & 23.0 & 22.9 & 22.8 \\
\rule{0pt}{3ex}Planck lensing & 8.8 & 8.8 & 8.7 & 8.5 \\
\rule{0pt}{3ex}BAO DESI DR2 & 12.4 & 12.2 & 11.4 & 9.7 \\
\rule{0pt}{3ex}SNe Pantheon+ & 1405.6 & 1404.2 & 1404.6 & 1403.0 \\
\hline
\rule{0pt}{3ex}\textbf{Total} & \textbf{4191.9} & \textbf{4187.6} & \textbf{4187.2} & \textbf{4183.2} \\
\rule{0pt}{3ex}$\Delta\chi^2$ & 0.0 & -4.3 & -4.7 & -8.7 
\end{tabular}
	\caption{\label{tab:MCMC}
Results for our combined analysis with  CMB (Planck 2018), SN (Pantheon+), and BAO (DESI DR2) data. Only if $95\,\%$ limits are one-sided, they are stated explicitly.\\$(\ast)${Unconstrained by Bayesian analysis within $95 \%$ CL. We refer in these cases to the profile likelihood analysis in Fig.~\ref{fig:DAO_profile}.}}
\end{table}

\end{appendix}
\bibliography{Ref}

@article{DESI:2025zgx,
    author = "Abdul Karim, M. and others",
    collaboration = "DESI",
    title = "{DESI DR2 results. II. Measurements of baryon acoustic oscillations and cosmological constraints}",
    eprint = "2503.14738",
    archivePrefix = "arXiv",
    primaryClass = "astro-ph.CO",
    reportNumber = "FERMILAB-PUB-25-0169-PPD",
    doi = "10.1103/tr6y-kpc6",
    journal = "Phys. Rev. D",
    volume = "112",
    number = "8",
    pages = "083515",
    year = "2025"
}

@misc{Garny:2026ish,
    author = "Garny, Mathias and Niedermann, Florian and Sloth, Martin S.",
    title = "{Dark Acoustic Oscillations and the Hubble Tension}",
    eprint = "2602.23895",
    archivePrefix = "arXiv",
    primaryClass = "astro-ph.CO",
    month = "2",
    year = "2026"
}

@article{Blas:2011rf,
    author = "Blas, Diego and Lesgourgues, Julien and Tram, Thomas",
    title = "{The Cosmic Linear Anisotropy Solving System (CLASS) II: Approximation schemes}",
    eprint = "1104.2933",
    archivePrefix = "arXiv",
    primaryClass = "astro-ph.CO",
    reportNumber = "CERN-PH-TH-2011-082, LAPTH-010-11",
    doi = "10.1088/1475-7516/2011/07/034",
    journal = "JCAP",
    volume = "07",
    pages = "034",
    year = "2011"
}

@article{Chen:2024tfp,
    author = "Chen, Shi-Fan and others",
    title = "{Baryon acoustic oscillation theory and modelling systematics for the DESI 2024 results}",
    eprint = "2402.14070",
    archivePrefix = "arXiv",
    primaryClass = "astro-ph.CO",
    doi = "10.1093/mnras/stae2090",
    journal = "Mon. Not. Roy. Astron. Soc.",
    volume = "534",
    number = "1",
    pages = "544--574",
    year = "2024"
}

@article{Schoneberg:2022grr,
    author = {Sch{\"o}neberg, Nils and Franco Abell{\'a}n, Guillermo},
    title = "{A step in the right direction? Analyzing the Wess Zumino Dark Radiation solution to the Hubble tension}",
    eprint = "2206.11276",
    archivePrefix = "arXiv",
    primaryClass = "astro-ph.CO",
    doi = "10.1088/1475-7516/2022/12/001",
    journal = "JCAP",
    volume = "12",
    pages = "001",
    year = "2022"
}

@article{Holm:2023uwa,
    author = "Holm, Emil Brinch and Nygaard, Andreas and Dakin, Jeppe and Hannestad, Steen and Tram, Thomas",
    title = "{PROSPECT: a profile likelihood code for frequentist cosmological parameter inference}",
    eprint = "2312.02972",
    archivePrefix = "arXiv",
    primaryClass = "astro-ph.CO",
    doi = "10.1093/mnras/stae2555",
    journal = "Mon. Not. Roy. Astron. Soc.",
    volume = "535",
    number = "4",
    pages = "3686--3699",
    year = "2024"
}

@article{Torrado:2020dgo,
    author = "Torrado, Jesus and Lewis, Antony",
    title = "{Cobaya: Code for Bayesian Analysis of hierarchical physical models}",
    eprint = "2005.05290",
    archivePrefix = "arXiv",
    primaryClass = "astro-ph.IM",
    reportNumber = "TTK-20-15",
    doi = "10.1088/1475-7516/2021/05/057",
    journal = "JCAP",
    volume = "05",
    pages = "057",
    year = "2021"
}

@article{Riess:2021jrx,
    author = "Riess, Adam G. and others",
    title = "{A Comprehensive Measurement of the Local Value of the Hubble Constant with 1 km s$^{−1}$ Mpc$^{−1}$ Uncertainty from the Hubble Space Telescope and the SH0ES Team}",
    eprint = "2112.04510",
    archivePrefix = "arXiv",
    primaryClass = "astro-ph.CO",
    doi = "10.3847/2041-8213/ac5c5b",
    journal = "Astrophys. J. Lett.",
    volume = "934",
    number = "1",
    pages = "L7",
    year = "2022"
}

@article{DESI:2025qqy,
    author = "Andrade, U. and others",
    collaboration = "DESI",
    title = "{Validation of the DESI DR2 measurements of baryon acoustic oscillations from galaxies and quasars}",
    eprint = "2503.14742",
    archivePrefix = "arXiv",
    primaryClass = "astro-ph.CO",
    reportNumber = "FERMILAB-PUB-25-0162-PPD",
    doi = "10.1103/kdys-w8vl",
    journal = "Phys. Rev. D",
    volume = "112",
    number = "8",
    pages = "083512",
    year = "2025"
}

@misc{Buen-Abad:2025bgd,
    author = "Buen-Abad, Manuel A. and Chacko, Zackaria and Flood, Ina and Kilic, Can and Marques-Tavares, Gustavo and Youn, Taewook",
    title = "{Dark Matter-Dark Radiation Interactions and the Hubble Tension}",
    eprint = "2511.16554",
    archivePrefix = "arXiv",
    primaryClass = "astro-ph.CO",
    reportNumber = "UT-WI-40-2025",
    month = "11",
    year = "2025"
}

@article{Bansal:2021dfh,
    author = "Bansal, Saurabh and Kim, Jeong Han and Kolda, Christopher and Low, Matthew and Tsai, Yuhsin",
    title = "{Mirror twin Higgs cosmology: constraints and a possible resolution to the H$_{0}$ and S$_{8}$ tensions}",
    eprint = "2110.04317",
    archivePrefix = "arXiv",
    primaryClass = "hep-ph",
    reportNumber = "FERMILAB-PUB-21-371-T",
    doi = "10.1007/JHEP05(2022)050",
    journal = "JHEP",
    volume = "05",
    pages = "050",
    year = "2022"
}

@article{Buen-Abad:2024tlb,
    author = "Buen-Abad, Manuel A. and Chacko, Zackaria and Flood, Ina and Kilic, Can and Marques-Tavares, Gustavo and Youn, Taewook",
    title = "{Atomic dark matter, interacting dark radiation, and the Hubble tension}",
    eprint = "2411.08097",
    archivePrefix = "arXiv",
    primaryClass = "hep-ph",
    reportNumber = "UTWI-37-2024",
    doi = "10.1007/JHEP07(2025)084",
    journal = "JHEP",
    volume = "07",
    pages = "084",
    year = "2025"
}

@article{Brout:2022vxf,
    author = "Brout, Dillon and others",
    title = "{The Pantheon+ Analysis: Cosmological Constraints}",
    eprint = "2202.04077",
    archivePrefix = "arXiv",
    primaryClass = "astro-ph.CO",
    doi = "10.3847/1538-4357/ac8e04",
    journal = "Astrophys. J.",
    volume = "938",
    number = "2",
    pages = "110",
    year = "2022"
}

@article{Planck:2018lbu,
    author = "Aghanim, N. and others",
    collaboration = "Planck",
    title = "{Planck 2018 results. VIII. Gravitational lensing}",
    eprint = "1807.06210",
    archivePrefix = "arXiv",
    primaryClass = "astro-ph.CO",
    doi = "10.1051/0004-6361/201833886",
    journal = "Astron. Astrophys.",
    volume = "641",
    pages = "A8",
    year = "2020"
}

@article{Planck:2018vyg,
    author = "Aghanim, N. and others",
    collaboration = "Planck",
    title = "{Planck 2018 results. VI. Cosmological parameters}",
    eprint = "1807.06209",
    archivePrefix = "arXiv",
    primaryClass = "astro-ph.CO",
    doi = "10.1051/0004-6361/201833910",
    journal = "Astron. Astrophys.",
    volume = "641",
    pages = "A6",
    year = "2020",
    note = "[Erratum: Astron.Astrophys. 652, C4 (2021)]"
}

@article{Vlah:2015zda,
    author = "Vlah, Zvonimir and Seljak, Uro{\v{s}} and Chu, Man Yat and Feng, Yu",
    title = "{Perturbation theory, effective field theory, and oscillations in the power spectrum}",
    eprint = "1509.02120",
    archivePrefix = "arXiv",
    primaryClass = "astro-ph.CO",
    doi = "10.1088/1475-7516/2016/03/057",
    journal = "JCAP",
    volume = "03",
    pages = "057",
    year = "2016"
}

@article{Blas:2016sfa,
    author = "Blas, Diego and Garny, Mathias and Ivanov, Mikhail M. and Sibiryakov, Sergey",
    title = "{Time-Sliced Perturbation Theory II: Baryon Acoustic Oscillations and Infrared Resummation}",
    eprint = "1605.02149",
    archivePrefix = "arXiv",
    primaryClass = "astro-ph.CO",
    reportNumber = "CERN-TH-2016-059, INR-TH-2016-009",
    doi = "10.1088/1475-7516/2016/07/028",
    journal = "JCAP",
    volume = "07",
    pages = "028",
    year = "2016"
}

@article{Senatore:2014via,
    author = "Senatore, Leonardo and Zaldarriaga, Matias",
    title = "{The IR-resummed Effective Field Theory of Large Scale Structures}",
    eprint = "1404.5954",
    archivePrefix = "arXiv",
    primaryClass = "astro-ph.CO",
    doi = "10.1088/1475-7516/2015/02/013",
    journal = "JCAP",
    volume = "02",
    pages = "013",
    year = "2015"
}

@article{Baldauf:2015xfa,
    author = "Baldauf, Tobias and Mirbabayi, Mehrdad and Simonovi{\'c}, Marko and Zaldarriaga, Matias",
    title = "{Equivalence Principle and the Baryon Acoustic Peak}",
    eprint = "1504.04366",
    archivePrefix = "arXiv",
    primaryClass = "astro-ph.CO",
    doi = "10.1103/PhysRevD.92.043514",
    journal = "Phys. Rev. D",
    volume = "92",
    number = "4",
    pages = "043514",
    year = "2015"
}

@article{Chen:2025mlf,
    author = "Chen, Shi-Fan and Zaldarriaga, Matias",
    title = "{It's all Ok: curvature in light of BAO from DESI DR2}",
    eprint = "2505.00659",
    archivePrefix = "arXiv",
    primaryClass = "astro-ph.CO",
    doi = "10.1088/1475-7516/2025/08/014",
    journal = "JCAP",
    volume = "08",
    pages = "014",
    year = "2025"
}

@misc{Sailer:2025lxj,
    author = "Sailer, Noah and Farren, Gerrit S. and Ferraro, Simone and White, Martin",
    title = "{Dispu$\tau$able: the high cost of a low optical depth}",
    eprint = "2504.16932",
    archivePrefix = "arXiv",
    primaryClass = "astro-ph.CO",
    month = "4",
    year = "2025"
}

@article{Beutler:2019ojk,
    author = "Beutler, Florian and Biagetti, Matteo and Green, Daniel and Slosar, An{\v{z}}e and Wallisch, Benjamin",
    title = "{Primordial Features from Linear to Nonlinear Scales}",
    eprint = "1906.08758",
    archivePrefix = "arXiv",
    primaryClass = "astro-ph.CO",
    doi = "10.1103/PhysRevResearch.1.033209",
    journal = "Phys. Rev. Res.",
    volume = "1",
    number = "3",
    pages = "033209",
    year = "2019"
}

@article{Seo:2015eyw,
    author = "Seo, Hee-Jong and Beutler, Florian and Ross, Ashley J. and Saito, Shun",
    title = "{Modeling the reconstructed BAO in Fourier space}",
    eprint = "1511.00663",
    archivePrefix = "arXiv",
    primaryClass = "astro-ph.CO",
    doi = "10.1093/mnras/stw1138",
    journal = "Mon. Not. Roy. Astron. Soc.",
    volume = "460",
    number = "3",
    pages = "2453--2471",
    year = "2016"
}

@article{Seo:2007ns,
    author = "Seo, Hee-Jong and Eisenstein, Daniel J.",
    title = "{Improved forecasts for the baryon acoustic oscillations and cosmological distance scale}",
    eprint = "astro-ph/0701079",
    archivePrefix = "arXiv",
    doi = "10.1086/519549",
    journal = "Astrophys. J.",
    volume = "665",
    pages = "14--24",
    year = "2007"
}

@article{Eisenstein:2006nj,
    author = "Eisenstein, Daniel J. and Seo, Hee-jong and White, Martin J.",
    title = "{On the Robustness of the Acoustic Scale in the Low-Redshift Clustering of Matter}",
    eprint = "astro-ph/0604361",
    archivePrefix = "arXiv",
    doi = "10.1086/518755",
    journal = "Astrophys. J.",
    volume = "664",
    pages = "660--674",
    year = "2007"
}

@article{Carlson:2012bu,
    author = "Carlson, Jordan and Reid, Beth and White, Martin",
    title = "{Convolution Lagrangian perturbation theory for biased tracers}",
    eprint = "1209.0780",
    archivePrefix = "arXiv",
    primaryClass = "astro-ph.CO",
    doi = "10.1093/mnras/sts457",
    journal = "Mon. Not. Roy. Astron. Soc.",
    volume = "429",
    pages = "1674",
    year = "2013"
}

@article{Matsubara:2007wj,
    author = "Matsubara, Takahiko",
    title = "{Resumming Cosmological Perturbations via the Lagrangian Picture: One-loop Results in Real Space and in Redshift Space}",
    eprint = "0711.2521",
    archivePrefix = "arXiv",
    primaryClass = "astro-ph",
    doi = "10.1103/PhysRevD.77.063530",
    journal = "Phys. Rev. D",
    volume = "77",
    pages = "063530",
    year = "2008"
}

@article{Matsubara:2008wx,
    author = "Matsubara, Takahiko",
    title = "{Nonlinear perturbation theory with halo bias and redshift-space distortions via the Lagrangian picture}",
    eprint = "0807.1733",
    archivePrefix = "arXiv",
    primaryClass = "astro-ph",
    doi = "10.1103/PhysRevD.78.109901",
    journal = "Phys. Rev. D",
    volume = "78",
    pages = "083519",
    year = "2008",
    note = "[Erratum: Phys.Rev.D 78, 109901 (2008)]"
}

@article{Bernal:2020pwq,
    author = "Bernal, Jos{\'e} Luis and Bellomo, Nicola and Raccanelli, Alvise and Verde, Licia",
    title = "{Beware of commonly used approximations. Part II. Estimating systematic biases in the best-fit parameters}",
    eprint = "2005.09666",
    archivePrefix = "arXiv",
    primaryClass = "astro-ph.CO",
    reportNumber = "CERN-TH-2020-054",
    doi = "10.1088/1475-7516/2020/10/017",
    journal = "JCAP",
    volume = "10",
    pages = "017",
    year = "2020"
}

@article{Tegmark:1997rp,
    author = "Tegmark, Max",
    title = "{Measuring cosmological parameters with galaxy surveys}",
    eprint = "astro-ph/9706198",
    archivePrefix = "arXiv",
    reportNumber = "IASSNS-AST-97-44",
    doi = "10.1103/PhysRevLett.79.3806",
    journal = "Phys. Rev. Lett.",
    volume = "79",
    pages = "3806--3809",
    year = "1997"
}

@article{Ivanov:2018gjr,
    author = "Ivanov, Mikhail M. and Sibiryakov, Sergey",
    title = "{Infrared Resummation for Biased Tracers in Redshift Space}",
    eprint = "1804.05080",
    archivePrefix = "arXiv",
    primaryClass = "astro-ph.CO",
    reportNumber = "CERN-TH-2018-076, INR-TH-2018-006",
    doi = "10.1088/1475-7516/2018/07/053",
    journal = "JCAP",
    volume = "07",
    pages = "053",
    year = "2018"
}

@article{Bayat:2025xfr,
    author = "Bayat, Zahra and Hertzberg, Mark P.",
    title = "{Examining quintessence models with DESI data}",
    eprint = "2505.18937",
    archivePrefix = "arXiv",
    primaryClass = "astro-ph.CO",
    doi = "10.1088/1475-7516/2025/08/065",
    journal = "JCAP",
    volume = "08",
    pages = "065",
    year = "2025"
}

@article{Khoury:2025txd,
    author = "Khoury, Justin and Lin, Meng-Xiang and Trodden, Mark",
    title = "{Apparent w{\ensuremath{<}}-1 and a Lower S8 from Dark Axion and Dark Baryons Interactions}",
    eprint = "2503.16415",
    archivePrefix = "arXiv",
    primaryClass = "astro-ph.CO",
    doi = "10.1103/w4qb-plk8",
    journal = "Phys. Rev. Lett.",
    volume = "135",
    number = "18",
    pages = "181001",
    year = "2025"
}

@misc{Kou:2025yfr,
    author = {Kou, Rapha{\"e}l and Lewis, Antony},
    title = "{Unified dark fluid with null sound speed as an alternative to phantom dark energy}",
    eprint = "2509.16155",
    archivePrefix = "arXiv",
    primaryClass = "astro-ph.CO",
    month = "9",
    year = "2025"
}

@misc{Chen:2025ywv,
    author = "Chen, Ruiqi and Cline, James M. and Muralidharan, Varun and Salewicz, Benjamin",
    title = "{Quintessential dark energy crossing the phantom divide}",
    eprint = "2508.19101",
    archivePrefix = "arXiv",
    primaryClass = "astro-ph.CO",
    month = "8",
    year = "2025"
}

@misc{Liu:2025bss,
    author = "Liu, Rayne and Zhu, Yijie and Hu, Wayne and Miranda, Vivian",
    title = "{Phantom Mirage from Axion Dark Energy}",
    eprint = "2510.14957",
    archivePrefix = "arXiv",
    primaryClass = "astro-ph.CO",
    month = "10",
    year = "2025"
}

@misc{Wang:2025znm,
    author = "Wang, Jia-Qi and Cai, Rong-Gen and Guo, Zong-Kuan and Wang, Shao-Jiang",
    title = "{Resolving the Planck-DESI tension by non-minimally coupled quintessence}",
    eprint = "2508.01759",
    archivePrefix = "arXiv",
    primaryClass = "astro-ph.CO",
    month = "8",
    year = "2025"
}

@misc{Caldwell:2025inn,
    author = "Caldwell, Robert R. and Linder, Eric V.",
    title = "{Null Impact of the Null Energy Condition in Current Cosmology}",
    eprint = "2511.07526",
    archivePrefix = "arXiv",
    primaryClass = "astro-ph.CO",
    month = "11",
    year = "2025"
}

@misc{SanchezLopez:2025uzw,
    author = "S{\'a}nchez L{\'o}pez, Samuel and Karam, Alexandros and Hazra, Dhiraj Kumar",
    title = "{Non-Minimally Coupled Quintessence in Light of DESI}",
    eprint = "2510.14941",
    archivePrefix = "arXiv",
    primaryClass = "astro-ph.CO",
    month = "10",
    year = "2025"
}

@misc{Bedroya:2025fwh,
    author = "Bedroya, Alek and Obied, Georges and Vafa, Cumrun and Wu, David H.",
    title = "{Evolving Dark Sector and the Dark Dimension Scenario}",
    eprint = "2507.03090",
    archivePrefix = "arXiv",
    primaryClass = "astro-ph.CO",
    month = "7",
    year = "2025"
}

@misc{Ferreira:2025lrd,
    author = "Ferreira, Elisa G. M. and McDonough, Evan and Balkenhol, Lennart and Kallosh, Renata and Knox, Lloyd and Linde, Andrei",
    title = "{The BAO-CMB Tension and Implications for Inflation}",
    eprint = "2507.12459",
    archivePrefix = "arXiv",
    primaryClass = "astro-ph.CO",
    month = "7",
    year = "2025"
}

@misc{H0DN:2025lyy,
    author = "Casertano, Stefano and others",
    collaboration = "H0DN",
    title = "{The Local Distance Network: a community consensus report on the measurement of the Hubble constant at 1{\%} precision}",
    eprint = "2510.23823",
    archivePrefix = "arXiv",
    primaryClass = "astro-ph.CO",
    month = "10",
    year = "2025"
}

@article{Knox:2019rjx,
    author = "Knox, Lloyd and Millea, Marius",
    title = "{Hubble constant hunter{\textquoteright}s guide}",
    eprint = "1908.03663",
    archivePrefix = "arXiv",
    primaryClass = "astro-ph.CO",
    doi = "10.1103/PhysRevD.101.043533",
    journal = "Phys. Rev. D",
    volume = "101",
    number = "4",
    pages = "043533",
    year = "2020"
}

@article{Bernal:2016gxb,
    author = "Bernal, Jose Luis and Verde, Licia and Riess, Adam G.",
    title = "{The trouble with $H_0$}",
    eprint = "1607.05617",
    archivePrefix = "arXiv",
    primaryClass = "astro-ph.CO",
    doi = "10.1088/1475-7516/2016/10/019",
    journal = "JCAP",
    volume = "10",
    pages = "019",
    year = "2016"
}

@article{Niedermann:2021vgd,
    author = "Niedermann, Florian and Sloth, Martin S.",
    title = "{Hot new early dark energy}",
    eprint = "2112.00770",
    archivePrefix = "arXiv",
    primaryClass = "hep-ph",
    doi = "10.1103/PhysRevD.105.063509",
    journal = "Phys. Rev. D",
    volume = "105",
    number = "6",
    pages = "063509",
    year = "2022"
}

@article{Niedermann:2021ijp,
    author = "Niedermann, Florian and Sloth, Martin S.",
    title = "{Hot new early dark energy: Towards a unified dark sector of neutrinos, dark energy and dark matter}",
    eprint = "2112.00759",
    archivePrefix = "arXiv",
    primaryClass = "hep-ph",
    doi = "10.1016/j.physletb.2022.137555",
    journal = "Phys. Lett. B",
    volume = "835",
    pages = "137555",
    year = "2022"
}

@article{Cruz:2023lnq,
    author = "Cruz, Juan S. and Niedermann, Florian and Sloth, Martin S.",
    title = "{NANOGrav meets Hot New Early Dark Energy and the origin of neutrino mass}",
    eprint = "2307.03091",
    archivePrefix = "arXiv",
    primaryClass = "astro-ph.CO",
    doi = "10.1016/j.physletb.2023.138202",
    journal = "Phys. Lett. B",
    volume = "846",
    pages = "138202",
    year = "2023"
}

@article{Garny:2024ums,
    author = "Garny, Mathias and Niedermann, Florian and Rubira, Henrique and Sloth, Martin S.",
    title = "{Hot new early dark energy bridging cosmic gaps: Supercooled phase transition reconciles stepped dark radiation solutions to the Hubble tension with BBN}",
    eprint = "2404.07256",
    archivePrefix = "arXiv",
    primaryClass = "astro-ph.CO",
    doi = "10.1103/PhysRevD.110.023531",
    journal = "Phys. Rev. D",
    volume = "110",
    number = "2",
    pages = "023531",
    year = "2024"
}

@misc{Garny:2025kqj,
    author = "Garny, Mathias and Niedermann, Florian and Rubira, Henrique and Sloth, Martin S.",
    title = "{Hot New Early Dark Energy: Dark Radiation Matter Decoupling}",
    eprint = "2508.03795",
    archivePrefix = "arXiv",
    primaryClass = "astro-ph.CO",
    month = "8",
    year = "2025"
}

@article{Poulin:2018cxd,
    author = "Poulin, Vivian and Smith, Tristan L. and Karwal, Tanvi and Kamionkowski, Marc",
    title = "{Early Dark Energy Can Resolve The Hubble Tension}",
    eprint = "1811.04083",
    archivePrefix = "arXiv",
    primaryClass = "astro-ph.CO",
    doi = "10.1103/PhysRevLett.122.221301",
    journal = "Phys. Rev. Lett.",
    volume = "122",
    number = "22",
    pages = "221301",
    year = "2019"
}

@article{Niedermann:2019olb,
    author = "Niedermann, Florian and Sloth, Martin S.",
    title = "{New early dark energy}",
    eprint = "1910.10739",
    archivePrefix = "arXiv",
    primaryClass = "astro-ph.CO",
    doi = "10.1103/PhysRevD.103.L041303",
    journal = "Phys. Rev. D",
    volume = "103",
    number = "4",
    pages = "L041303",
    year = "2021"
}

@article{Niedermann:2020dwg,
    author = "Niedermann, Florian and Sloth, Martin S.",
    title = "{Resolving the Hubble tension with new early dark energy}",
    eprint = "2006.06686",
    archivePrefix = "arXiv",
    primaryClass = "astro-ph.CO",
    doi = "10.1103/PhysRevD.102.063527",
    journal = "Phys. Rev. D",
    volume = "102",
    number = "6",
    pages = "063527",
    year = "2020"
}

@article{Cruz:2023lmn,
    author = "Cruz, Juan S. and Niedermann, Florian and Sloth, Martin S.",
    title = "{Cold New Early Dark Energy pulls the trigger on the H $_{0}$ and S $_{8}$ tensions: a simultaneous solution to both tensions without new ingredients}",
    eprint = "2305.08895",
    archivePrefix = "arXiv",
    primaryClass = "astro-ph.CO",
    doi = "10.1088/1475-7516/2023/11/033",
    journal = "JCAP",
    volume = "11",
    pages = "033",
    year = "2023"
}

@article{Schoneberg:2024ynd,
    author = {Sch{\"o}neberg, Nils and Vacher, L{\'e}o},
    title = "{The mass effect {\textemdash} variations of the electron mass and their impact on cosmology}",
    eprint = "2407.16845",
    archivePrefix = "arXiv",
    primaryClass = "astro-ph.CO",
    doi = "10.1088/1475-7516/2025/03/004",
    journal = "JCAP",
    volume = "03",
    pages = "004",
    year = "2025"
}

@article{Poulin:2024ken,
    author = "Poulin, Vivian and Smith, Tristan L. and Calder{\'o}n, Rodrigo and Simon, Th{\'e}o",
    title = "{Implications of the cosmic calibration tension beyond H0 and the synergy between early- and late-time new physics}",
    eprint = "2407.18292",
    archivePrefix = "arXiv",
    primaryClass = "astro-ph.CO",
    doi = "10.1103/PhysRevD.111.083552",
    journal = "Phys. Rev. D",
    volume = "111",
    number = "8",
    pages = "083552",
    year = "2025"
}

@article{Jeong:2013eza,
    author = "Jeong, Kwang Sik and Takahashi, Fuminobu",
    title = "{Self-interacting Dark Radiation}",
    eprint = "1305.6521",
    archivePrefix = "arXiv",
    primaryClass = "hep-ph",
    reportNumber = "TU-938, IPMU13-0114",
    doi = "10.1016/j.physletb.2013.07.001",
    journal = "Phys. Lett. B",
    volume = "725",
    pages = "134",
    year = "2013"
}

@article{Buen-Abad:2015ova,
    author = "Buen-Abad, Manuel A. and Marques-Tavares, Gustavo and Schmaltz, Martin",
    title = "{Non-Abelian dark matter and dark radiation}",
    eprint = "1505.03542",
    archivePrefix = "arXiv",
    primaryClass = "hep-ph",
    doi = "10.1103/PhysRevD.92.023531",
    journal = "Phys. Rev. D",
    volume = "92",
    number = "2",
    pages = "023531",
    year = "2015"
}

@article{Buen-Abad:2017gxg,
    author = "Buen-Abad, Manuel A. and Schmaltz, Martin and Lesgourgues, Julien and Brinckmann, Thejs",
    title = "{Interacting Dark Sector and Precision Cosmology}",
    eprint = "1708.09406",
    archivePrefix = "arXiv",
    primaryClass = "astro-ph.CO",
    doi = "10.1088/1475-7516/2018/01/008",
    journal = "JCAP",
    volume = "01",
    pages = "008",
    year = "2018"
}

@article{Archidiacono:2020yey,
    author = "Archidiacono, Maria and Gariazzo, Stefano and Giunti, Carlo and Hannestad, Steen and Tram, Thomas",
    title = "{Sterile neutrino self-interactions: $H_0$ tension and short-baseline anomalies}",
    eprint = "2006.12885",
    archivePrefix = "arXiv",
    primaryClass = "astro-ph.CO",
    doi = "10.1088/1475-7516/2020/12/029",
    journal = "JCAP",
    volume = "12",
    pages = "029",
    year = "2020"
}

@article{Blinov:2020hmc,
    author = "Blinov, Nikita and Marques-Tavares, Gustavo",
    title = "{Interacting radiation after Planck and its implications for the Hubble Tension}",
    eprint = "2003.08387",
    archivePrefix = "arXiv",
    primaryClass = "astro-ph.CO",
    reportNumber = "FERMILAB-PUB-20-114-A-T",
    doi = "10.1088/1475-7516/2020/09/029",
    journal = "JCAP",
    volume = "09",
    pages = "029",
    year = "2020"
}

@article{Aloni:2021eaq,
    author = "Aloni, Daniel and Berlin, Asher and Joseph, Melissa and Schmaltz, Martin and Weiner, Neal",
    title = "{A Step in understanding the Hubble tension}",
    eprint = "2111.00014",
    archivePrefix = "arXiv",
    primaryClass = "astro-ph.CO",
    doi = "10.1103/PhysRevD.105.123516",
    journal = "Phys. Rev. D",
    volume = "105",
    number = "12",
    pages = "123516",
    year = "2022"
}

@article{Kaplan:2009de,
    author = "Kaplan, David E. and Krnjaic, Gordan Z. and Rehermann, Keith R. and Wells, Christopher M.",
    title = "{Atomic Dark Matter}",
    eprint = "0909.0753",
    archivePrefix = "arXiv",
    primaryClass = "hep-ph",
    doi = "10.1088/1475-7516/2010/05/021",
    journal = "JCAP",
    volume = "05",
    pages = "021",
    year = "2010"
}

@article{Cyr-Racine:2012tfp,
    author = "Cyr-Racine, Francis-Yan and Sigurdson, Kris",
    title = "{Cosmology of atomic dark matter}",
    eprint = "1209.5752",
    archivePrefix = "arXiv",
    primaryClass = "astro-ph.CO",
    doi = "10.1103/PhysRevD.87.103515",
    journal = "Phys. Rev. D",
    volume = "87",
    number = "10",
    pages = "103515",
    year = "2013"
}

@article{Cyr-Racine:2021oal,
    author = "Cyr-Racine, Francis-Yan and Ge, Fei and Knox, Lloyd",
    title = "{Symmetry of Cosmological Observables, a Mirror World Dark Sector, and the Hubble Constant}",
    eprint = "2107.13000",
    archivePrefix = "arXiv",
    primaryClass = "astro-ph.CO",
    doi = "10.1103/PhysRevLett.128.201301",
    journal = "Phys. Rev. Lett.",
    volume = "128",
    number = "20",
    pages = "201301",
    year = "2022"
}

@article{Blinov:2021mdk,
    author = "Blinov, Nikita and Krnjaic, Gordan and Li, Shirley Weishi",
    title = "{Toward a realistic model of dark atoms to resolve the Hubble tension}",
    eprint = "2108.11386",
    archivePrefix = "arXiv",
    primaryClass = "hep-ph",
    reportNumber = "FERMILAB-PUB-21-344-T, FERMILAB-PUB-21-344-T",
    doi = "10.1103/PhysRevD.105.095005",
    journal = "Phys. Rev. D",
    volume = "105",
    number = "9",
    pages = "095005",
    year = "2022"
}

@article{Bansal:2022qbi,
    author = "Bansal, Saurabh and Barron, Jared and Curtin, David and Tsai, Yuhsin",
    title = "{Precision cosmological constraints on atomic dark matter}",
    eprint = "2212.02487",
    archivePrefix = "arXiv",
    primaryClass = "hep-ph",
    doi = "10.1007/JHEP10(2023)095",
    journal = "JHEP",
    volume = "10",
    pages = "095",
    year = "2023"
}
\end{document}